\documentclass[%
 aip,
 amsmath,amssymb,
 reprint,%
]{revtex4-1}

\usepackage{graphicx}
\usepackage{dcolumn}
\usepackage{bm}

\usepackage[utf8]{inputenc}
\usepackage[T1]{fontenc}
\usepackage{mathptmx}

\begin{document}

\preprint{AIP/123-QED}

\title{Affine Quantization of the Harmonic Oscillator on the Semi-bounded domain $(-b,\infty)$ for $b: 0 \rightarrow \infty$\\}

\author{Carlos R. Handy}
\email{carlos.handy@tsu.edu}
 \affiliation{Department of Physics, Texas Southern University, 3100 Cleburne St., Houston, TX 77004}
 

\date{\today}

\begin{abstract}
The transformation of a classical system into its quantum counterpart is usually done through the well known procedure of canonical quantization.  However, on non-Cartesian domains, or on bounded  Cartesian domains, this procedure can be plagued with theoretical inconsistencies. An alternative approach is  {\it affine quantization} (AQ) Fantoni and Klauder (Phys. Rev. D {\bf 103}, 076013 (2021)), resulting in different conjugate variables that lead to a more consistent quantization formalism.  To highlight these issues, we examine a deceptively simple, but important, problem: that of the harmonic oscillator potential on the semibounded domain: ${\cal D} = (-b,\infty)$. The AQ version of this corresponds to the (rescaled) system, ${\cal H} = \frac{1}{2}\Big(-\partial_x^2 + \frac{3}{4(x+b)^2} + x^2\Big)$. We solve this system numerically for $b > 0$. The case $b = 0$ corresponds to an {\it exactly solvable} potential, for which the eigenenergies can be determined exactly (through non-wavefunction dependent methods), confirming the results of Gouba ( J. High Energy Phys., Gravitation Cosmol. {\bf 7}, 352-365 (2021)). We investigate the limit $b \rightarrow \infty$, confirming that the full harmonic oscillator problem is recovered. The adopted computational methods are in keeping with the underlying theoretical framework of AQ. Specifically, one method is an affine map invariant variational procedure, made possible through a moment problem quantization reformulation. The other method focuses on boundedness (i.e. $L^2$) as an explicit quantization criteria. Both methods lead to converging bounds to the discrete state energies; and thus confirming the accuracy of our results, particularly as applied to a singular potential problem.
\end{abstract}

\maketitle

\section{Introduction}
In a recent series of works, Klauder and other researchers$^{1-8}$ have advocated for the relevance of {\it affine quantization} (AQ) as a remedy  to the limitations of {\it canonical quantization} (CQ). These methods have been shown to be particularly effective in the quantization of gravity and other systems, involving formulations devoid of non-renormalizability problems, etc.

The central motivation is that for  systems on nonlinear, or constrained Cartesian, domains, {\it canonical quantization} fails, as amply reviewed by Gouba$^9$.

These concepts can be demonstrated through quantum mechanics. To this extent, there has been recent interest in understanding one dimensional quantum systems on semi-bounded domains from the perspective of {\it affine quantization}$^9$. 

To better understand these issues we focus on a deceptively simple system. Consider the {\it walled} harmonic oscillator problem, corresponding to a quantum particle on the entire real axis, $\Re$, but restricted to the ${\cal D}_b \equiv (-b,\infty)$ domain by an infinite barrier potential on $x \leq -b$, and an harmonic oscillator potential for $x > -b$. This problem is well defined; with an Hermitian Hamiltonian and a Hermitian linear momentum, as defined through canonical quantization (CQ), on the $L^2$ space of functions for the entire real domain, $L^2(\Re,dx)$. 

Now consider applying CQ to a quantum particle defined solely on the semi-bounded domain, ${\cal D}_b$, under the influence of the corresponding harmonic oscillator potential. Any attempt at a consistent CQ formulation, on the full space of $L^2$ functions on the corresponding ${\cal D}_b$ domain, denoted by $L^2({\cal D}_b,dx)$, is impossible. 

The importance of {\it affine quantization} (AQ) is that, unlike CQ, it does offer a consistent quantization formulation on $L^2({\cal D}_b,dx)$.

In this work we make explicit all the above issues. In addition to numerically studying the AQ formulation for $b \geq 0$ we also show that the full harmonic oscillator is recovered in the $b \rightarrow \infty$ limit.

The AQ version of the aforementioned quantum system  is the Hamiltonian    ${\cal H} = \frac{1}{2}\Big(-{\hbar}^2\partial_\chi^2 + \frac{3}{4}\chi^{-2}+(\chi-b)^2\Big)$, for $x \equiv \chi-b \geq -b $, as derived below. The behavior of this singular system, as $b\rightarrow \infty $ has not been studied. We do so through two formalism that resonate with the theoretical framework of AQ,  leading to two powerful computational methods, each resulting in rapidly converging lower and upper bounds to the discrete state energies.
\vspace{-2em}

\subsection {Positivity, Affine Transformations and the Moment Problem}
Positivity and affine transformations (i.e. dilation operator) are essential components of AQ. Coincidentally, the first computational method to be used in this work  is based on a Moment Problem$^{10}$ reformulation of the underlying Sturm-Liouville problem. The relevant method, the Eigenvalue Moment Method (EMM)$^{11-16}$, corresponds to an affine map invariant variational procedure over the space of polynomials, exploiting  the positivity of the associated physical configurations. It leads to the generation of geometrically converging lower and upper bounds to the discrete quantum states amenable to this type of representation. Given the singular nature of the above referenced potential, the availability of tight bounds leads to highly accurate results.

For the (large and important) class of problems amenable to this type of moment representation analysis, the system is transformed into one with an exact algebraic structure, to all orders (i.e. no truncated approximations are involved). Furthermore, the use of an extensive representation, as opposed to the local representation of configuration space, makes the use of a power moment representation more efficient in solving the eigenenergy problem.

\subsection{Boundedness as a Quantization Criterion}

Usually {\it boundedness} (i.e. working in $L^2$ spaces) is implicit in standard quantization representations. However, the moment formulations used here explicitly work within a representation in which the physical (i.e. $L^2$) and unphysical states are simultaneously present, and one must define quantization criteria that are unique to the bounded states. 

Related to this is the second major theme of {\it affine quantization}: the importance of defining Hermitian operators that act on the full $L^2$ space of functions for the constrained domain in question. We shall denote this by $L^2({\cal D},dx)$. On constrained Cartesian domains, Canonical Quantization produces Hamiltonians that  are non-self adjoint on the full $L^2({\cal D},dx)$ space of wavefunctions. Therefore, {\it boundedness} (i.e. working on the full space of bounded, $L^2({\cal D},dx)$ configurations) becomes an explicit quantization criterion within AQ. In the following discussion, any reference to $L^2$ will be with respect to the space of square integrable functions on the domain in question.

Our second, and more robust, computational method also generates tight bounds, but uses $L^2$ {\it boundedness} as an explicit quantization criterion. This second computational method is referred to as the Orthonormal Polynomial Projection Quantization  Bounding Method (OPPQ-BM) method$^{17}$, exploiting the use of (non-orthonormal) weighted polynomial basis expansions as introduced by Handy and Vrinceanu$^{18,19}$.

The OPPQ formulation, like EMM, is also based on transforming the given (local) configuration space system into an extensive representation defined through a (power) moments reformulation of the given configuration space system. That is, $\Psi \rightarrow \{\mu(p)|p \geq 0\}$ where $ \mu(p) = \int dx \ x^p \Psi(x)$, and which are required to satisfy a moment equation relation (MER) corresponding to a (multidimensional) linear, homogeneous,  finite difference equation.

\subsection{The Relevance of Moment Representations In Multiscale Analysis}

We note another important feature of our moment representation based computational methods and its relevance to the very nature of {\it affine quantization}.

Strong coupling physics is usually associated with singular perturbation analysis involving kinetic energy perturbations (i.e. $\frac{\hbar^2}{2m} \rightarrow 0$)$^{20}$. 
These involve understanding how systems adapt to regions of abrupt changes (i.e. on constrained Cartesian domains defined by  boundaries, edges, boundary layers, etc.). Moment based quantization was introduced as a multiscale analytical framework to regulate such singular perturbation-strong coupling expansions$^{21}$, which arise not only in solving strong coupling problems$^{12,17}$, but also semiclassical expansions (i.e. $\hbar \rightarrow 0$), and  large mass expansions,
(i.e. $m \rightarrow \infty$). 

Indeed, the power moments of a system, $\mu(p) \equiv \int dx \ x^p\Psi(x)$,  define the inverse scale (i.e. $\frac{1}{a}$) expansion of the scaling transform$^{21}$, 
${S\Psi}(a,b) \equiv \frac{1}{a\nu}\int dx \ S\big(\frac{x-b}{a}\big)\Psi(x)$, 
which is the structure that couples the large and small scale features of a system (i.e. $a:\infty \rightarrow 0$). 

It is well known that wavelets, one of the most efficient multi-resolution representations $^{22,23}$, is derived from the scaling transform representation $^{24}$. Wavelets define an efficient resummation procedure for the inverse scale expansion of the scaling transform, as defined by the power moments. As such, the methods introduced here are in keeping with the challenges of CQ, and the remedy afforded by AQ.

Below we develop the essential perspectives for understanding the limitations of CQ, and the advantages of AQ. We also review, in greater detail, the issues previously introduced. 

\section {Affine Quantization: Correcting the Limitations of Canonical Quantization}

Canonical quantization (CQ) has become the standard approach for quantizing classical systems. For one dimensional systems, for simplicity, the CQ procedure identifies the conjugate canonical variables, $\{q,p\}$, and the corresponding classical Hamiltonian, ${\cal H}(p,q)$. These are then transformed through the Schr\"odinger representation, $q \rightarrow  Q \equiv x$ and $p \rightarrow P \equiv \frac{\hbar}{i}\partial_x$,  resulting in an Hermitian operator expression ${\cal H}(x,\frac{\hbar}{i}\partial_x)$, after careful attention to the ordering of the $Q$ and $P$ operators.

For the remainder of this discussion, we will assume that the physical parameters have been rescaled to unity: $\hbar \rightarrow 1$, etc.

 According to Klauder$^{3}$ and Gouba$^9$, CQ only works when the quantum particle(s) are defined on unconstrained Cartesian domains. However, this  approach can be plagued with inconsistencies for systems on  curved or otherwise constrained domains. 
 
Consider a quantum system whose classical counterpart is restricted to the positive coordinate, $x > 0$. We then want the position operator to be positive definite, $Q > 0$, on the space  $L^2(\Re^+,dx)$, corresponding to the space of $L^2$ functions on the positive real axis. When this happens, the usual conjugate momentum operator, $P$, cannot be self adjoint, since in general the elements of $L^2(\Re^+,dx)$ include functions which are not zero at the origin, $\Psi(0) \neq 0$. That is, $\langle \Psi|\frac{1}{i}\partial_x\Phi\rangle = i\Psi^*(0)\Phi(0) +\langle\frac{1}{i}\partial_x \Psi|\Phi\rangle$. 

One could relax the positivity condition, by working on the entire real axis. Thus, consider the Schr\"odinger equation problem:
\begin{eqnarray}
-\partial_x^2\Psi(x) + V(x)\Psi(x) = 2E \Psi(x),
\end{eqnarray}
for the barrier-harmonic oscillator potential
\begin{eqnarray}
V(x) = \begin{cases}V_0\ >\ 0 , \ for\ x < 0, \\
 x^2,\  x\ \geq 0.
\end{cases}
\end{eqnarray}
The Hamiltonian acts on the space $L^2(\Re,dx)$, and the solutions to Eq.(1) are $L^2$ configurations with finitely discontinuous second derivatives. 

Needless to say, the position operator $Q = x$ is no longer positive definite, and the conjugate canonical momentum, $P = \frac{1}{i}\partial_x$, is self adjoint, since all $L^2$ functions must vanish at $x = \pm \infty$.

The $L^2$ eigenfunctions to Eqs.(1-2), on the positive domain (i.e. $\Re^+$), exist for arbitrary $E > 0$. Through continuity conditions on the wavefunction and its first derivative we know that  the bounded configurations on the entire real axis (i.e. $2E < V_0$, and $\Psi(x) = {\cal N} e^{-\sqrt{V_0-2E}  |x|}$, for $x < 0$),  must satisfy $\frac{\Psi(0^+)}{\Psi'(0^+)} \equiv {\cal E}(E)= \frac{1}{\sqrt{V_0-2E}}$, where ${\cal E}(E)$ is a determinable function of the energy. This is the quantization condition for finite $V_0 < \infty$.

When $V_0 \rightarrow \infty$, the eigenenergies for bound states on the entire real axis, $\Re$, correspond to those values that give  ${\cal E}(E) = \frac{\Psi(0^+)}{\Psi'(0^+)} = 0$.  Since $\Psi(x)$ is a solution to the harmonic oscillator potential, for $x > 0$, it and its derivative cannot both be zero at the origin.  Given that  $\Psi(0) = {\cal N}$ and $\Psi'(0) = {\cal N}\sqrt{V_0-E}$,  we see that we can satisfy the quantization condition by demanding that ${\cal N} \rightarrow 0$ and $ {\cal N}\sqrt{V_0-E} \rightarrow finite$, as $V_0 \rightarrow \infty$. This then leads to the usual quantization condition for the corresponding infinite barrier potential problem (i.e. {\it Walled Harmonic Oscillator}): $\Psi(0^+) = 0$, with eigenenergies $\{\frac{3}{2},\frac{7}{2},\frac{11}{2},\ldots\}$.

The {\it walled harmonic oscillator} (WHO) corresponds to the infinite  barrier-harmonic oscillator potential ($V_0 \rightarrow \infty$)
\begin{eqnarray}
V_{WHO}(x) = \begin{cases} \infty,\ x < 0 ,\\
x^2,\ x\geq 0.
\end{cases}
\end{eqnarray}
When viewed in the context of $L^2(\Re,dx)$ (i.e. the  $V_0 \rightarrow \infty$ limit of the finite barrier-harmonic oscillator potentials), the WHO system's eigenstates are well defined, and correspond to
\begin{eqnarray}
\Psi(x) =\begin{cases} 0, \ x \leq 0 \\
{\cal O}_\eta(x),\ x \geq 0,
\end{cases}
\end{eqnarray}
where ${\cal O}_\eta(x) = {\cal N}_\eta H_{2\eta+1}(x) exp(-\frac{x^2}{2})$, involving the odd order hermite polynomials. We will refer to these as the {\it odd} states.

The configurations, ${\cal E}_{\eta} \equiv {\cal N}_\eta H_{2\eta}(x) exp(-\frac{x^2}{2})$ despite being solutions to the ${\cal H}_{WHO}$ Hamiltonian operator, on the positive domain, do not satisfy the required boundary condition. Indeed, they play no role within the spectrum of bound states for the finite step potentials, and in particular, the infinite barrier potential. One might ascribe the latter  to the derivative being discontinuous at the origin, leading to an infinite energy. 

\subsection{The Half Harmonic Oscillator}

When viewed as a problem on $L^2(\Re^+,dx)$, then the  {\it walled harmonic oscillator} becomes the {\it half harmonic oscillator}. Relative to this new space, it is not Hermitian (from the perspective of CQ); although it is Hermitian over two subspaces:
\begin{eqnarray}
S_{\cal E} & \equiv & \{\Psi(x)| \partial_x\Psi(0) = 0, \Psi\in L^2(\Re^{+},dx)\} , \\
S_{\cal O} & \equiv & \{ \Psi(x)| \Psi(0) = 0, \Psi\in L^2(\Re^{+},dx)\} .
\end{eqnarray}
Each of these subspaces contains square integrable eigenstates. Unlike the full $\Re$ formulation, the subspace ${S_{\cal E}}$ is not filtered out within the CQ formulation.

Adding to the confusion is the fact that the $S_{\cal E}$ is dense with respect to the {\it physical subspace}, $S_{\cal O}$. This is trivial to prove, as argued, and demonstrated, in the Appendix. Thus, the eigenstates within $S_{\cal O}$, with eigenvalues $\{\frac{3}{2},\frac{7}{2},\frac{11}{2},\ldots\}$ can be regarded as the (infinite) superposition of the eigenstates of $S_{\cal E}$ with eigenvalues $\{\frac{1}{2},\frac{5}{2},\frac{9}{2},\ldots\}$. This is the inconsistency of canonical quantization.

The {\it momentum operator} $-i\partial_x$ is Hermitian only over the $S_{\cal O}$ subspace. 
We could explicitly restrict the physical space to $S_{\cal O}$. What we want is an alternate formalism that introduces a Hamiltonian that essentially does this automatically. This is, in effect, what {\it affine quantization} does.


\subsection{An Hermitian - Resolution: Affine Quantization}
 
 To address the previous dilemma concerning the {\it half harmonic oscillator} problem on the positive real axis, $\Re^{+}$,  Klauder has advocated for {\it affine quantization} (AQ). It involves keeping a positive position operator, $Q > 0$, but identifying a different conjugate variable for the canonical momentum. This alternative choice is self-adjoint on $L^2(\Re^+,dx)$. 
 
As noted, the basic problem is that on the space $L^2(\Re^+,dx)$, the conventional, canonical quantization (CQ), momentum operator is not self-adjoint,  $P \neq P^\dag$. An alternative choice begins with the dilation operator: $D\equiv  (PQ+QP)/2$, where $P = \frac{1}{i}\partial_x$, and $Q = x = Q^\dagger$. It then follows that $D^\dagger = D$. This is confirmed by explicitly using the differential forms for $Q$ and $P$.

An alternative derivation follows from $P^\dag Q = P Q$. If we define $\langle \Phi|P^\dag Q|\Psi\rangle \equiv \langle P \Phi|Q\Psi\rangle$, then it follows (through substituting the differential forms for $P$ and $Q$) that $ \langle P \Phi|Q\Psi\rangle = \langle  \Phi|PQ\Psi\rangle$; or, $P^\dag Q = P Q$. Taking $D \equiv P^\dag Q+QP$, makes the self-adjointness, $D^\dag = D$, evident at the pure operator level. We can now also express the dilation operator as: $D =  ( PQ+QP )/2$.

A different way of appreciating the operator  $D$ follows from canonical quantization: $[Q,P]=i\hbar 1\!\!1$. Upon multiplying this expression by $Q$ we obtain $i\hbar Q=(Q[Q,P]+[Q,P]Q)/2$, which becomes
$[Q,D]=i\hbar\,Q$. If $Q$ is invertible (i.e. $Q > 0$ or $Q < 0$) then we have $[Q,Q^{-1}D] = i\hbar 1\!\!1$, as well as $[Q,DQ^{-1}] = i\hbar 1\!\!1$.
It then follows that  $({Q^{-1}D)^\dagger} = D Q^{-1}$, and $DQ^{-2}D$ becomes self-adjoint. This will become the new kinetic energy operator.

Having $Q>0$ (i.e. facilitating its invertibility) is just the rule that can help the half-harmonic oscillator.  

\subsection{Affine Quantization's Configuration Space Representation for the Half-Harmonic Oscillator}

The classical harmonic oscillator Hamiltonian, once again, is $H(p,q)=(p^2+q^2)/2$, after a rescaling that introduces an overall factor of $\frac{1}{2}$. Using $d\equiv pq$, gives us for the classical hamiltonian,  ${\tilde H}(d,q)= (d^2/q^2+q^2)/2$. The affine quantization version of this  is
\begin{eqnarray}
{\cal H}_A =(DQ^{-2} D +Q^2)/2.
\end{eqnarray}
Adopting Schr\"odinger's representation leads to $Q=x>0$ and 
 $D= -i\hbar [ x\frac{d}{dx} +\frac{d}{dx}x ]/2=-i\hbar [x \frac{d}{dx}+1/2]$. 
 
 The affine quantization of the harmonic oscillator involves the Hamiltonian:
 \begin{eqnarray}
     {\cal H} & = &\frac{1}{2}\Big( -\hbar^2 [x \frac{d}{dx}+1/2] x^{-2} [x \frac{d}{dx}+1/2] + x^2\Big)  \nonumber \\
          & = &\frac{1}{2}\Big( -\hbar^2 \frac{d^2}{dx^2} + (3/4) \hbar^2 \frac{1}{x^2} + x^2 \Big),
\end{eqnarray}
  which is referred to as          
            a {\it spiked harmonic oscillator}$^{25}$.
            
            This differential eigenenergy problem ${\cal H}\Psi = E \Psi$,
            was  recently studied by L. Gouba$^9$, who observed
            that it has 
              eigenvalues of $2\hbar (n+1)$ for $ n =0,1,2,3,...$. These energy levels are  equally spaced and 
             twice the spacing of the full-harmonic oscillator, or $E_n = \hbar(n+\frac{1}{2})$. Indeed, this system corresponds to an {\it exactly solvable} system$^{26,27}$, and we can recover these exact energies through a different method (wavefunction independent), as outlined in the following sections. 
             
             We can also consider more general, constrained Cartesian domains, such as ${\cal D}_b \equiv (-b,\infty)$, for $b \geq 0$. In this case, we work with $\chi = x+b \geq 0$:
          
          \begin{eqnarray}
     {\cal H}_{A;b} = \frac{1}{2}\Big( -\hbar^2 \frac{d^2}{d\chi^2} + (3/4) \hbar^2\frac{1}{\chi^2} + (\chi-b)^2 \Big).
\end{eqnarray} 

             An important question is, will the limit $b \rightarrow \infty$ recover the spectra of the full harmonic oscillator. One expects the answer to be in the affirmative. However, the exact manner in which this happens has not been previously solved, numerically. This is the primary objective of this work. 
             
             The CQ counterpart to the $b\rightarrow \infty$ limit, since it is a formulation on the entire real axis, is referred to as {\it the wall sliding away}. The AQ formulation has no {\it wall} as such, since it is solely defined on the positive real axis.
             
\section {The Relevance of the Adopted Computational Formalism}

One of the central themes of {\it affine quantization} (AQ) is the use of the dilation operator as an alternate canonical variable. This is a generator for affine transformations over the coordinate space. Another important theme is that AQ focuses on developing a consistent quantization formulation over the entire $L^2$ space on which a system is defined.

We will study the eigenenergy structure of the AQ Hamiltonian in Eq.(9) through two different computational procedures that are consistent with the underlying theortical framework of affine quantization. Each is capable of generating tight bounds to the discrete states of the system. 

Before elaborating upon the above, we note that the computational procedures studied here have one common feature. They all depend on the quantum system being transformable into a configuration representation, $\Psi \rightarrow {\cal F}$, whose Stieltjes power moments,

\begin{eqnarray}
v(p) \equiv \int_0^\infty \ d\chi \ \chi^p {\cal F}(\chi). 
\end{eqnarray}
satisfy a linear, homogeneous, finite difference equation of order $1+m_s$, as defined below.   This is referred to as a moment equation representation (MER) relation. These MER considerations apply for problems in one, or several, space dimensions.

The MER relation takes on the generator form:
\begin{eqnarray}
\nu(p) = \sum_{\ell = 0}^{m_s} M_E(p,\ell) \ \nu_\ell,
\end{eqnarray}
$p \geq 0$,
in which a subset of the power moments, referred to as the {\it missing moments}, $\{\nu_\ell| \nu(\ell) \equiv \nu_\ell,\ 0 \leq \ell \leq m_s\}$, generate all the other moments through known, energy dependent coefficients. The missing moments are required to satisfy some appropriate normalization condition.

\subsection{The Eigenvalue Moment Method: An Affine Map Invariant, Variational, Bounding Method}

The first computational method is referred to as the Eigenvalue Moment Method (EMM)$^{11-16}$. It involves transforming Eq.(9) into a MER representation for ${\cal F}_1 \equiv \Psi$ or ${\cal F}_2 \equiv \Psi^2$ (i.e. including ${\cal F}_3 \equiv \Psi^*\Psi$, for non-Hermitian systems), and then imposing positivity constraints arising from the associated Moment Problem theorems$^{10}$. These constraints are associated with the Hankel-Hadamard (HH) determinantal, positivity, constraints. They in turn correspond (in their quadratic form formulation) to an affine map invariant, variational  analysis over the space of polynomials. However, the space of polynomials is invariant under affine transformations; therefore, EMM is an affine map invariaint variational procedure for generating tight bounds to the discrete states.

The computational implementation of EMM requires the direct, or indirect, use of nonlinear convex optimization$^{28,29}$. This is now referred to as semidefinite programming (SDP). It can be relaxed through linear programming$^{30}$, the original computational implementation of EMM.  The first use of  SDP related analysis  to solve quantum operators (i.e. partial differential equations) was by Handy and Bessis$^{11,12}$, as acknowledged by Lasserre$^{29}$. Until the works by Handy and Bessis, it had never occurred to anyone to combine the MER representation, for suitable systems, with the positivity properties of certain physical solutions (in configuration space), and apply the Moment Problem theorems to constrain the energy parameter (through geometrically converging lower and upper bounds).

It is within the EMM representation that we can generate the exact energies for the $b = 0$ case corresponding to the {\it spiked harmonic oscillator} in Eq.(9). This analysis, given in a subsequent section, requires no explicit reference to the wavefunctions. 

Our existing codes are Fortran based (i.e. 14 digits of precision) and we can only implement EMM to a limited moment expansion order. Mathematica has only recently implemented SDP analysis in its codes, permitting, in principle, much higher precision analysis. Nevertheless, our limited EMM results serve as a benchmark for a purely algebraic bounding analysis, that of our second computational procedure: the Orthonormal Polynomial Projection Quantization - Bounding Method (OPPQ-BM)$^{17}$. The {\it weighted polynomial} basis expansion of this approach is modeled after earlier works by Handy and Vrinceanu$^{18,19}$. The latter is referred to in this work as the OPPQ-Approximation Method since the underlying algebraic properties of the formalism (made possible by the MER representation) allow for the generation of high accuracy (i.e. rapidly converging) approximations to the eigenenergies. OPPQ-AM is faster to implement than OPPQ-BM, although the latter is used to confirm the results. Both are used in this work.

\subsection {Improving the EMM Convergence}
The best way to improve the convergence rate of the EMM bounds is to transform the system into a configuration space representation involving fewer missing moments. Usually, this can be accomplished through a well chosen contact transformation, ${\cal F}(\chi) = {\cal R}(\chi) \Psi(\chi)$. We usually find that if the positive configuration $ {\cal R}(\chi) > 0$ is modeled after the asymptotic form of the ground state (or the asymptotic form of the discrete states), the missing moment order within the new representation is reduced, or even made to be zero. 

Not all $m_s = 0$ MER representations correspond to {\it exactly solvable systems}. One example is the sextic anharmonic oscillator, $V(x) = mx^2 + gx^6$ (i.e. ${\cal R}(x) = exp(-\frac{\sqrt{g}x^4}{4})$). However, if a system is {\it exactly solvable} the MER relation for an appropriate contact transformation (i.e. usually the actual ground state), will reveal the exact energies. We implement this in the corresponding section below, confirming not only the exact form for the energies, but also the exact form of the wavefunctions, confirming Gouba's (more traditional) analysis$^9$. 

\subsection {The Orthonormal Polynomial Projection Quantization Method}

We briefly describe the second computational approach pursued in this work. It involves exploiting the 
 same MER relationship  within the $\Psi$ representation, combined with the explicit use of $L^2$ boundedness to determine the eigenenergies. This also leads to an eigenenergy bounding strategy that is purely algebraic, and implementable to arbitrary order through Mathematica. The unlimited accuracy of Mathematica allows us to work at high precision, to arbitrary order. 
 
 This second, eigenenergy bounding method, is built on the  Orthonormal Polynomial Projection Quantization (OPPQ) formalism, previously introduced by Handy and Vrinceanu$^{18,19}$. Their ansatz led to a very effective approximation method (AM) for the discrete state energies. We refer to it, in this work, as OPPQ-AM. Its reformulation as a bounding method (BM) was only recently discovered by Handy$^{17}$, leading to impressive bounds for low lying excited states of the quadratic Zeeman problem for superstrong magnetic fields. These bounds are consistent with, or surpass, the most accurate estimates in the literature as given  by Kravchenko et al$^{31}$, and Schimerczek and Wunner$^{32}$. It can also be extended to multidimensional non-hermitian systems (for which OPPQ-AM is also effective) , as well as multidimensional bosonic and fermionic systems.

The essentials of the OPPQ analysis involve expanding $\Psi$ in terms of a non-orthogonal basis, ${\cal B}_j(\chi) \equiv P_j(\chi)R(\chi)$, involving the orthonormal polynomials relative to the positive weight, $R(\chi) > 0$:
\begin{eqnarray}
\Psi(\chi) = \sum_{j=0}^\infty c_j \ P_j(\chi) R(\chi),
\end{eqnarray}
where 
\begin{eqnarray}
\langle P_{j_1}|R|P_{j_2}\rangle =\delta_{j_1,j_2}, 
\end{eqnarray}
(note $\langle {\cal B}_{j_1}|{\cal B}_{j_2}\rangle \neq \delta_{j_1,j_2}$).
One can then generate the projection coefficients, exactly, through the underlying MER relation
\begin{align}
c_j(E;{\overrightarrow \nu})& = \langle P_j|\Psi\rangle, \nonumber \\
& = \sum_{\ell=0}^{m_s} \Lambda_\ell^{(j)}(E)\ \nu_\ell,
\end{align}
where ${\overrightarrow \nu} \equiv (\nu_0,\ldots,\nu_{m_s})$, and the energy dependent coefficients are known in closed form, through the underlying MER relation. For one dimensional systems, we can adopt a unit missing moment vector normalization, $|{\overrightarrow \nu}| = 1$.

Define the expression:
\begin{eqnarray}
{\cal I}[\Psi,R]\equiv \int_0^\infty d\chi  \frac{\Psi^2(\chi)}{R(\chi)}.
\end{eqnarray}
As long as the weight, $R(\chi) > 0$, is chosen so that it does not decrease faster than the asymptotic form of the physical solutions (i.e. either as $\chi \rightarrow \infty$ or $\chi \rightarrow 0^+$), then the integral is finite for physical configurations, and infinite for unphysical configurations. This is an explicit use of $L^2$ boundedness as a quantization criterion.

It follows that upon substituting the OPPQ expansion in Eq.(12) we obtain 
\begin{eqnarray}
{\cal I}[\Psi,R] = {\cal I}[E,{\overrightarrow \nu}] = \sum_{j=0}^\infty c_j^2(E;{\overrightarrow \nu}).
\end{eqnarray}

To quantize, we require that this integral expression be finite for the exact physical energies and missing moments; and infinite, otherwise:
\begin{eqnarray}
{\cal I}[E,{\overrightarrow \nu}] = \begin{cases}
finite \iff E = E_{phys}\ {\rm and} \ {\overrightarrow \nu} = {\overrightarrow \nu}_{phys}; \\
\infty \iff E \neq E_{phys}\ {\rm or} \ {\overrightarrow \nu} \neq {\overrightarrow \nu}_{phys}.
\end{cases}
\end{eqnarray}
The major focus of the OPPQ-BM formalism is to reduce Eq.(17) into a minimization problem in the energy parameter leading to the generation of bounds. The details are given here, as applied to the system in Eq.(9).



\section{Essentials of the EMM Bounding Method}

In this section we examine the EMM bounding procedure on three different configuration space representations. The first is EMM-$\Psi$ which will yield bounds for the ground state, since it is the only physical configuration that is positive on $\Re^+$.

The second application extends EMM to the configuration ${\cal F}(\chi) = {\cal R}(\chi) \Psi(\chi)$, where the ${\cal R} > 0$ expression is modeled after the asymptotic form of the physical states. Enhanced convergence of the ground state energy bounds is achieved through a reduction in the missing moment order. However, for the case of $b = 0$, this procedure transforms the problem into one corresponding to an {\it exactly solvable} system and all the discrete state energies are recovered exactly. 

The third EMM application is within the $\Psi^2$ representation. We are able to generate bounds to the low lying discrete states. 

All of the above results are used to confirm the high accuracy results made possible through the alternative bounding analysis, that of OPPQ-BM, as discussed in the following section.

\subsection{EMM-$\Psi$: Bounding the Ground State Energy}

 We examine the EMM formalism as applied to Hamiltonian in Eq.(9), where all physical parameters have been rescaled to unity (i.e. $\hbar \rightarrow 1$, etc.). We also have regrouped certain terms  obtaining:

\begin{equation}
- \partial_\chi^2\Psi + \Big(\frac{3}{4} \frac{1}{\chi^2} + (\chi^2+\beta \chi)\Big)\Psi= \lambda \Psi,
\end{equation}
where $\beta \equiv -2 b$ and $\lambda \equiv 2E -b^2$. The origin corresponds to a regular singular point, generating independent solutions of the form
\begin{eqnarray}
Y_j(\chi) = \begin{cases} \chi^{\frac{3}{2}} A_1(\chi), \ for \ j = 1 , \\
\chi^{\frac{3}{2}} Ln (\chi) A_1(\chi) + \chi^{-{\frac{1}{2}}} A_2(\chi), \ for \ j = 2,
\end{cases}
\end{eqnarray}
where the $A_j$ are analytic near the origin. The power series expansion for $A_1$ is uniquely determined, and it in turn generates $A_2$. 

Our immediate objective is to generate a moment equation for the physical solutions to Eq.(18). 
Any solution to Eq.(18) must be a linear superposition of these two independent configurations, $\Psi(\chi) =c_1 Y_1(\chi)+c_2 Y_2(\chi)$. The physical, $L^2$ solutions, must asymptotically vanish at infinity according to the zeroth order WKB expression $\Psi(\chi) \rightarrow exp(-{\frac{1}{2}}\chi^2)$. 
The $L^2$ condition, $\int_0^\infty \Psi^2(\chi) < \infty$,  leads to the following conditions for the physical solutions:

\begin{eqnarray}
\Psi_{phys}(\chi) \sim \begin{cases} \chi^{\frac{3}{2}} A_1(\chi),\  \chi \rightarrow 0^+ , \\
{\cal N} exp(-{\frac{1}{2}}\chi^2), \ \chi \rightarrow \infty.
\end{cases}
\end{eqnarray}

Define the power moments as:
\begin{eqnarray}
v(p) \equiv \int_0^\infty d\chi \ \chi^p \Psi(\chi).
\end{eqnarray}
There will be unphysical solutions that also decay at infinity, 
$\Psi_{unphys}(\chi) \sim {\cal N} exp(-{\frac{1}{2}}\chi^2)$, as $\chi \rightarrow \infty$, but behave near the origin according to $\Psi_{unphys}(\chi) = Y_2(\chi)$, as given Eq.(19). These unphysical solutions will have finite power moments for
\begin{eqnarray}
 |v_{unphys}(p)| < \infty, \ if \ p > -{\frac{1}{2}}.\cr
\end{eqnarray}
By way of contrast, the physical power moments will be finite so long as
\begin{eqnarray}
 |v_{phys}(p)| < \infty, \ if 
 \ p >  -\frac{5}{2}.\cr
\end{eqnarray}
The correct MER relation must involve moment orders  starting within the range 
\begin{eqnarray}
-\frac{5}{2} < p_{initial}\leq -\frac{1}{2}.
\end{eqnarray}
If not, then both physical and unphysical solutions will satisfy the EMM relation and no bounds will be generated.

Upon multiplying both sides of Eq.(18) by $\chi^p$ and integrating by parts, incorporating the $Y_1(\chi)$ behavior near the origin, and insuring no boundary terms appear (although not necessary, it will complicate the EMM formalism if we include boundary terms), we obtain
 the MER for the physical solutions :
\begin{eqnarray}
 v(p+2)=   -\beta v(p+1)   + \lambda v(p) + \Big(p(p-1) -\frac{3}{4}  \Big)v(p-2),\cr
\end{eqnarray}
$p > -1/2$.  Additionally, we also require that
\begin{eqnarray}
-1/2 < p_{initial} \leq 3/2 ,
\end{eqnarray}
in order to filter out unphysical solutions that are exponentially bounded at infinity, but become unbounded near the origin.

To further clarify things, we note that the lowest order term in Eq.(25) is the $\nu(p-2)$ moment. We require that the initial $p$ values fall within the range: $-5/2 < p-2 \leq -1/2$, in order that Eq.(24) is satisfied. This means that $-1/2 < p_{initial} \leq 3/2$. The natural choices are $p_{initial} = 0, \frac{1}{2}, 1, \frac{3}{2}$. Clearly, the weakest option is to take $p_{intial} = 3/2$, since then the MER relation will only imply the physical solution at that $p$-value. The strongest option is $p = 0$, since now the MER relation involves four moment constraints unique to the physical solutions.

The EMM formalism is easier to implement if we work with the nonnegative integer power moments of an appropriate configuration. Working with the $v(p)$ moments in Eq.(25) requires that we incorporate the $v(-1)$ and $v(-2)$ power moments. To avoid modifying the EMM formalism (i.e. to accomodate negative integer order power moments), we simply work with a slightly modified configuration. To this extent, define the power moments for the physical solutions: 
\begin{eqnarray}
u(p) \equiv  \int_0^\infty d\chi \chi^{p} \chi^{-2}\Psi_{phys}(\chi),
\end{eqnarray}
for $p \geq 0$. If we take $v(p) \equiv u(p+2)$,
then the corresponding $u$-moment equation becomes:
\begin{align}
 u(p+4) \ =   -\beta u(p+3) & +  
 \lambda u(p+2) \nonumber \\ & + \Big(p(p-1) -\frac{3}{4}  \Big)u(p),\cr
\end{align}
$p > -\frac{1}{2}$, although we will restrict $p = integer \geq 0$. Note that Eq.(28) is manifestly an $m_s = 3$ missing moment order problem (i.e. assuming $p \geq 0$). 

For $b = 0$, or $\beta = 0$, Eq.(28) separates into two lower order moment equations for the even and odd order power moments. We can implement EMM on these; however, we do not quote the results here since they are inferior to simply working with Eq.(28) directly.

The results of implementing EMM on the $\Psi$ representation is given in Table 1, for various values for $b= 0, 10, 20,  \ldots$. The tightness of the eigenenergy bounds for the ground state are limited by the low machine precision (i.e. 14 digits) of our Fortran code, and low moment expansion order indicated $P_{max} < O(30)$. Nevertheless, these results provide a useful guide.

\begin{table}
\caption{
$EMM-\Psi_{gr}$.
}
\centerline{
\begin{tabular}{rrrr}
\hline
$b$ & $E^{(L)}$    &  $E^{(U)}$  &   $P_{max}$   \\
\hline
 0& 1.999415&  2.000489&  29 \\
.1 & 1.870371 & 1.871507& 28 \\
.5 & 1.428965 & 1.429646 & 30\\
1 & 1.032844 & 1.033323 & 28 \\
5 & 0.515648 & 0.516333 & 23  \\
10 & 0.496295 & 0.524709 & 16  \\
20 & 0.457391&  0.658807 & 13  \\
 \hline
\hline
\end{tabular}}
\end{table}

\subsection{EMM in the $\Phi(x) = exp(-\frac{x^2}{2}) \Psi(x) $ Representation}

An alternative EMM strategy for improving the tightness of the bounds is to choose a representation consistent with the EMM theory but involving fewer missing moments. One way to do this is to work with a contact transformation incorporating the asymptotic form (at infinity) of the physical solutions: 
\begin{eqnarray}
\Phi(x) = exp(-\frac{x^2}{2}) \Psi(x).
\end{eqnarray}
This will generally result in a MER relation with reduced missing moment order. 

Note that when altering the configuration space through a contact transformation, it is important that physical(unphysical) solutions in the $\Psi$ representation transform into physical(unphysical) configurations in the $\Phi$ representation:
\begin{eqnarray}
\Psi_{phys/unphys} \rightarrow \Phi_{phys/unphys}.
\end{eqnarray}
That is, we want physical $\Psi$ configurations with finite power moments, and unphysical $\Psi$ configurations with infinite power moments, to transform into the new representation, $\Psi \rightarrow \Phi$, while preserving these properties.  

We can transform the Schrodinger equation into a differential equation for 
$\Phi(\chi) =
exp\big( -\frac{1}{2}(\chi-b)^2\big)\Psi(\chi)$:

\begin{eqnarray}
-\Big(\Phi''(\chi) + 2 (\chi-b)\Phi'(\chi)\Big) + \frac{3}{4}\frac{1}{\chi^2}\Phi(\chi) = (2E+1) \Phi(\chi),\cr
\end{eqnarray}
$\chi \geq 0$.
Since the $\Phi$ representation differs from the $\Psi$ representation by an analytic factor, the local structure (i.e. $\chi \approx 0$) of the $\Phi$ solutions is identical to that of the $\Psi$'s. 
Implementing the same analysis as before, upon multiplying both sides by $\chi^{p}$ and integrating by parts, being mindful of avoiding boundary terms, and exploiting the $\chi \approx 0$ form of the $\Phi$ solutions (identical to that in Eq.(19), we obtain the MER for the power moments $v(p) = \int_0^\infty d\chi\ \chi^p \Phi(\chi)$:

\begin{eqnarray}
 v(p) = \frac{(p+1/2)(p-3/2)  v(p-2)+2bpv(p-1)} {(2p+1-2E)},\cr
\end{eqnarray}
for $p > -1/2$.  As before, the initial $p$-value must satisfy $-1/2 < p_{initial} \leq 3/2$, since initiating the recursion relation at higher values would not filter out the unphysical solutions that are exponentially decaying at infinity. 

The MER relation in Eq.(32) is of a lower missing moment order than that of Eq.(25). Essentially, one goes from $m_s = 3$ (i.e. Eq.(25)  requires $\nu(p-2),\nu(p-1),\nu(p),\nu(p+1)$ to generate $\nu(p+2)$) to $m_s = 1$ (i.e. Eq.(32) requires $\nu(p-2),\nu(p-1)$ to generate $\nu(p)$). The computational results of this representation are described after discussing the $b = 0$ case in the following subsection.

\subsubsection {The Exact Eigenenergies for the $b = 0$ Case}

One can verify that for $b = 0$ (i.e. $\chi = x$) the exact ground state wavefunction, up to a normalization, is $\Psi_{gr}(\chi) = \chi^\frac{3}{2} exp(-\frac{1}{2} \chi^2)$. This motivates the following transformation.

Let us work with ${\tilde v}(p) \equiv v(p+3/2)$. We then obtain from Eq.(32):

\begin{eqnarray}
{\tilde v}(p) = \frac{(p+2)p  {\tilde v}(p-2)+2b(p+3/2){\tilde v}(p-1)}{2(p+2-E)},\cr
\end{eqnarray}
for $1 < p_{initial}+3/2 \leq  3$. Thus, we can take $p \geq 0$ in Eq.(33).

For the case $b = 0$ this MER relation dramatically simplifies: 

\begin{eqnarray}
{\tilde v}(p) = \frac{p(p+2)  {\tilde v}(p-2)}{2(p+2-E)},\cr
\end{eqnarray}
for $p \geq 0$. The first observation is that 
\begin{eqnarray}
(2-E){\tilde v}(0) = 0.
\end{eqnarray}
We know that the ground state must be nonnegative, on the nonnegative real axis, and therefore all its power moments must be positive. This then tells us that

\begin{eqnarray}
E_{gr} \equiv E_0 = 2.
\end{eqnarray}
For the excited states, ${\tilde v}_{exc}(0) = 0$. 

If we now take $p = 2$ in Eq.(34) we conclude that
\begin{eqnarray}
(4-E){\tilde v}(2) = 4{\tilde v}(0);
\end{eqnarray}
however, all the excited states must have a zeroth order moment that is zero. Accordingly, $E_{1st} \equiv E_1 = 4$, and ${\tilde v}_{n \geq 2}(2) = 0$, where $n$ corresponds to the quantum number (i.e. $n = 0,1,2,\ldots$).

For $p = 4$ we conclude that $2(6-E){\tilde \nu}(4) = 24{\tilde \nu}(2)$. We then conclude that $E_{2nd} = 6$, and ${\tilde\nu}_{n \geq 3}(4) = 0$.
In this manner, one can readily argue that
\begin{eqnarray}
E_n = 2(n+1).
\end{eqnarray}
Additionally, for the quantum number $n = 0,1,\ldots$:
\begin{eqnarray}
{\tilde v}_{n}(2q)  =  0, \ if \  q \leq n-1, n \geq 1.
\end{eqnarray}
We note that the even order power moments satisfy
\begin{eqnarray}
{\tilde v}(2q) & = & v(2q+3/2) = \int_0^\infty d\chi \ \chi^{2(q+\frac{3}{4})}\Phi(\chi), \nonumber \\
& = & \frac{1}{2} \int_0^\infty d\xi \ \xi^{(q+\frac{1}{4})}\Phi(\sqrt{\xi}) \nonumber \\
& \equiv & w(q), \ q\geq 0,
\end{eqnarray}
where $\xi \equiv \chi^2$.

The $\{w(q)\}$ are the Stieltjes power moments of the function $\Upsilon(\xi) \equiv \frac{1}{2}\xi^{\frac{1}{4}}\Phi(\sqrt{\xi})$. The ground state, in this representation, corresponds to $\Upsilon_0(\xi) =\frac{1}{2} \xi^{\frac{1}{4}} exp(-\frac{1}{2}\xi) \times \xi^{\frac{3}{4}} exp(-\frac{1}{2}\xi) = \frac{\xi}{2}\ e^{-\xi} $. Now assume that the (unnormalized) excited states correspond to $\Upsilon_n(\xi) =  {\cal P}_n(\xi) \Upsilon_0(\xi)$, where ${\cal P}_n(\xi)$ are the orthogonal polynomials relative to the ground state, $\Upsilon_0(\xi)$, as a weight. That is $\langle {\cal P}_n|\Upsilon_0|{\cal P}_m\rangle = {\cal N}_m\delta_{m,n}$.  We see that the behavior of the power moments in Eq.(39) is identical to that of the power moments of such  orthogonal polynomial expressions. That is $\langle \xi^q|\Upsilon_n\rangle = 0$, for $q \leq n-1$, provided $n \geq 1$. 

Exactly solvable systems generally involve discrete state wavefunctions that are the orthogonal polynomials relative to one of the classic weights. In this case, the orthogonal polynomials of $\xi exp(-\xi)$ correspond to the Associated Laguerre polynomials ${\cal P}_n(\xi) = {\cal L}_n^{(1)}(\xi)$. Although we have determined the exact physical energies, without requiring explicit knowledge of the wavefunctions; to validate the above conjectured form of the physical states requires some simple analysis. 

First of all, the Associated Laguerre polynomials $y(\xi) \equiv {\cal L}_n^{(1)}(\xi)$,  satisfy the equation 
$\xi y''(\xi)+(2-\xi)y'(\xi)+n y(\xi) = 0$. Taking $\Upsilon(\xi)  = y(\xi) \xi exp(-\xi)$, give us the differential equation
\begin{eqnarray}
\xi \Upsilon''(\xi) + \xi \Upsilon'(\xi) + (n+1) \Upsilon(\xi) = 0.
\end{eqnarray}
Let $\omega(q)$ correspond to the Stieltjes moments. We then have
the corresponding MER:
\begin{eqnarray}
q(q+1) \omega(q-1)-(q+1)\omega(q) +(n+1)\omega(q) = 0,
\end{eqnarray}
$q \geq 0$.
This gives the recursion relation
\begin{eqnarray}
\omega(q) = \frac{q(q+1) \omega(q-1)}{q-n},
\end{eqnarray}
for $q \geq 0$. This is the same moment equation satisfied by the $w(q)$ moments in Eq.(40), or from Eq.(34), $w(q) = \frac{q(q+1) w(q-1)}{q+1-(n+1)}$.

Working backwards from the $\Upsilon$ representation to the $\Psi$ representation we obtain:

\begin{eqnarray}
\Upsilon_n(\xi) & = & {\cal P}_n(\xi) \Upsilon_0(\xi), \nonumber \\
\Phi_n(\chi) & = & {\cal P}_n(\chi^2) \Phi_0(\chi),\nonumber \\
\Psi_n(\chi) & = & {\cal P}_n(\chi^2) \Psi_0(\chi), \nonumber \\
\Psi_n(\chi) & = & {\cal P}_n(\chi^2)\times \chi^{3/2} \ exp(-\chi^2/2),
\end{eqnarray}
where ${\cal P}_n(\xi) = {\cal L}_n^{(1)}(\xi)$, corresponding to the Associated Laguerre polynomials.

That is, within the $\chi$ representation we see that the eigenstates correspond to the even power orthogonal polynomials for the square of the ground state, $\Psi_0^2(\chi) = \chi^3 exp(-\chi^2)$, or $\int_0^\infty d\chi \ P_m(\chi^2)P_n(\chi^2) \Psi_0^2(\chi) = 2 \delta_{m,n}$. They are the ordinary orthonormal polynomials in the $\xi \equiv \chi^2$ representation for the weight $\Upsilon_0(\xi) = \frac{1}{\sqrt{\xi}} \Psi_0^2(\sqrt{\xi}) = \xi exp(-\xi)$.

\subsubsection{Algebraic Bounds for the Ground State, $b \geq 0$}

Returning to Eq.(32), for $p = 0$, and $b \neq 0$, since the ground state must have all its power moments be positive and finite, it follows that 
\begin{eqnarray}
E_{gr}(b) > {1/2}, \ b \geq 0;
\end{eqnarray}
whereas upon taking $p = 3/2$ we obtain
\begin{eqnarray}
E_{gr}(b) < 2, \ b  > 0.
\end{eqnarray}

\subsubsection {Numerical Bounds for the Ground State, $b > 0$}

In Eq.(32) we have $-1/2 < p_{initial} \leq 3/2$. We want to make this selection more explicit. If $p = q +\frac{\sigma}{2}$, where  $q \geq 0$, then the condition on the initial $p$-index becomes: $-1 < \sigma \leq 3$.

Define $u(q) \equiv v(q+{\frac{\sigma}{2}}-2)  = \int_0^\infty d\chi \  \chi^{ q+{\frac{\sigma}{2}}-2} \Phi(\chi)$, the power moments of the $\chi^{ \frac{\sigma}{2}-2} \Phi(\chi)$ configuration.  We then obtain the MER relation:

\begin{flalign}
 u( & q+2)  = \nonumber \\
 &= \frac
 {\Big({\frac{3}{4}}-(q+\frac{\sigma} {2})(q+\frac{\sigma} {2}-1)\Big) u(q)   -2b\big( q+{\frac{\sigma} {2}} \big)\ u(q+1)}
  {\Big(2E-1-2(q+{\frac{\sigma} {2}}) \Big)}  ,\cr
 \end{flalign}
$q\geq 0$, and $-1 < \sigma \leq 3$.

These MER relations are of missing moment order $m_s = 1$; however, for $\sigma =3$, the missing moment order is actually zero, $m_s = 0$.

For $\sigma = 3$ they are effectively of order $m_s = 0$, since the $u(0)$ moment does not contribute. That is, only the moments $\{u(q)|q \geq 1\}$ couple. In this case, one prefers to work with the moments:
\begin{eqnarray}
 \tau(q)\equiv u(q+1),
 \end{eqnarray}
 for $q \geq 0$. We then obtain

\begin{eqnarray}
 \Big(2E-4\Big) \tau(1)  =  -3b \ \tau(0), 
\end{eqnarray}
where we can set $\tau(0) \equiv 1$; and
\begin{eqnarray}
 \tau(q+1)  = \frac{q(q+2)\tau(q-1) 
  +2b\big( q+\frac{3}{2} \big) \tau(q)}{\Big(4+2q-2E\Big)}, \cr 
\end{eqnarray}
for $q \geq 1$. 

Both formulations (i.e. $\sigma = 0,3$) dramatically reduce the missing moment order, as compared to the $\Psi$ formulation. The improved convergence, or tightness, of the bounds can be seen in Table 2.

We note that the $\sigma = 0$ bounds are superior for $O(b) > 5$; whereas the $\sigma = 3$ bounds are superior for $O(b) < 5$. We believe that part of the reason is that the $\sigma = 0$ formulation contains two moment constraints that are unique  to the bound states (i.e. $q = 0,1$); whereas the $\sigma = 3$ contains only one such relation (i.e. that for $q = 0$).

\begin{table}
\caption{\label{tab:table1}
$EMM-\Phi_{gr}, m_s = 1/0\ (\sigma = 0/3), Eqs.(47-50)$.
}
\begin{ruledtabular}
\begin{tabular}{rllrr}
\hline 
$b$ & $E^{(L)}$    &  $E^{(U)}$  &   $P_{max}$  & $\sigma$  \\
\hline
0 &  1.999714   &2.000244 & 27 & $0$ \\
& 2  & 2$^*$ & 1 & $3$\\
\hline
.001&    1.9986710464441 & 1.9986710464498$^*$  &    16  & $3$\\
\hline
$.01$ & 1.9867452618193 & 1.9867452618204$^*$  &    20  &  $3$\\
\hline
.1 & 1.870636 & 1.871151& 27 & $0$ \\
 & 1.8709141846102 & 1.8709141846107$^*$ &  22 & $ 3$ \\
  \hline
.5 & 1.429056 & 1.429492 & 30 & $0$ \\
& 1.4292927197475 & 1.4292927197522$^*$& 24 &  $3$\\
\hline
1 & 1.032928& 1.033250 & 27 & $0$ \\
&1.0331033239001 &   1.0331033239766$^*$ & 25 &   $3$\\
\hline
5 & 0.51598078 & 0.51598081 & 18 & $0$ \\
   & 0.51591386 & 0.51598114$^*$ &   20 & $3$\\
 \hline
10&0.5038074052 & 0.5038074090& 13 &$0$ \\
20 & 0.5009410333&  0.5009410338 & 10& $0$ \\
100 &0.5000375056&0.5000375257& 7 & $0$ \\
1000&0.500000375000431&0.500000375004431& 7 & $0$ \\
2500 & 0.500000059999800 & 0.500000060003038 & 6 & $0$ \\
 \hline
* & Eq.[47] with $\sigma =3$ & are Eqs.[49,50] with  $m_s = 0 $  &  & \\
\hline

\end{tabular}
\end{ruledtabular}
\end{table}

\subsection{The EMM-$\Psi^2$ Moment Equation Representation}

The previous formulations were applicable only to the ground state. An alternate strategy for obtaining bounds to the low lying discrete states is to work with the probability density, $S(\chi) \equiv \Psi^2(\chi)$. 

The probability density for the Schrodinger equation, $-\Psi''(x) + V(x) \Psi(x) = {\cal E}\Psi(x)$, with real potential, $V(x)$,  satisfies a third order linear, ordinary, differential equation$^{14,15}$ (i.e. if $V$ is complex, then $\Psi^*(x)\Psi(x)$ satisfies a fourth order LODE$^{16}$):

\begin{eqnarray}
-S'''(x) + 4( V(x) - {\cal E}) S'(x)  + 2V'(x) S(x) =0.
\end{eqnarray}
In the present case, where we have $-\Psi''+(\frac{3}{4}(x+b)^{-2}+x^2)\Psi(x) = 2E\Psi(x)$, with ${\cal E} = 2E$, we obtain:
\begin{flalign}
 -S'''(\chi)+\Big( {3}\chi^{-2}+ & 4 (\chi-b)^2-8E\Big) S'(\chi)   \nonumber \\
 &+ 2\Big( 2(\chi-b)-\frac{3}{2}\chi^{-3}  \Big) S(\chi) = 0.\cr
\end{flalign}

The three fundamental solutions to Eq.(52) will be (refer to Eq.(19))

\begin{flalign}
\Lambda_j(\chi)& = \\
&\begin{cases}  Y_1^2(\chi)  = \chi^3 {\cal A}_1(\chi), \ j = 1, \nonumber \\
Y_1(\chi) Y_2(\chi) =\chi^3Ln(\chi){\cal A}_1(\chi)+\chi{\cal A}_2(\chi) , \ j = 2, \nonumber \\
Y_2^2(\chi) ={
\begin{cases}\chi^3Ln^2(\chi){\cal A}_1(\chi)+  2\chi Ln(\chi) {\cal A}_2(\chi) \nonumber \\  + \chi^{-1}{\cal A}_3(\chi), \ j = 3
\end{cases}} .
\end{cases}
\cr
\end{flalign}
where ${\cal A}_1(\chi) = A_1^2(\chi)$, ${\cal A}_2(\chi) = A_1(\chi) A_2(\chi)$, and ${\cal A}_3(\chi) = A_2^2(\chi)$. 

The physical solution must be of the form $\Lambda_1(\chi)$. This will have finite Stieltjes moments $v(p) = \int_0^\infty d\chi \ \chi^p \Lambda_1(\chi)$, as long as  $p >-4$. Proceeding with a similar analysis to that given in the two previous subsections, the moment equation for the physical $S(\chi) \equiv \Psi^2(\chi)$ solutions becomes

\begin{eqnarray}
 4(1+p)v(1+p) =  
4b(1+2p)v(p)+(8E-4b^2)pv(p-1) \nonumber \\ 
+(p-1)\Big(p(p-2)-3\Big)v(p-3), \nonumber \cr
\end{eqnarray}
$p \geq 0$, or more generally $p > -1$. 

Let us define $v(p-3) \equiv u(p)$, it then follows that:
\begin{flalign}
4(1+p) u(p+4) =&\  4b(1+2p)u(p+3) \nonumber \\
&+(8E-4b^2)pu(p+2) \nonumber \\
&+(p-1)\Big(p(p-2)-3\Big)u(p),  \cr
\end{flalign}
$p\geq 0$.

\begin{table}
\caption{
$EMM-\Psi^2, Eq.(54)$.
}
\centerline{
\begin{tabular}{rrllr}
\hline
$b$ & State &$E^{(L)}$    &  $E^{(U)}$  &   $P_{max}$   \\
\hline
0 & Gr &1.99998825   &2.00005739 & 26\\
0 & $1^{st}$&3.99671544 & 4.00447362 & 25 \\
0& $2^{nd}$&5.91768300 & 6.04294214 & 26\\
\hline
.1 & Gr &1.87090202 & 1.87092049& 26 \\
.1 & $1^{st}$&3.82074273 & 3.82158281 & 28 \\
.1&$2^{nd}$&     5.75422040 & 5.82182702 & 27\\
\hline
.5 & Gr&1.42928910 & 1.42929704& 30 \\
.5& $1^{st}$&3.18388603&3.18432342&28\\
.5&$2^{nd}$&4.97456267&5.02884966&27\\
\hline
1 & Gr &          1.03310195& 1.03310458 & 29 \\
1 & $1^{st}$ &2.55658870& 2.55901222 & 27 \\
1 & $2^{nd}$&4.12209511 &4.19851998& 27 \\
\hline
5 & Gr &0.51278794 & 0.52143406 & 22 \\
5&$1^{st}$ &1.475 & 1.855 & 23 \\
 \hline
\hline
\end{tabular}}
\end{table}

The EMM-$\Psi^2$ results are given in Table 3. Based on the previous experiences, for $b = 0$ we ignore the separability of the even and odd order $u(p)$ moments, and treat the system as an $m_s = 3$ missing moment order system. The results for the ground state are consistent with, although inferior to,  those in Table 2 for the EMM-$\Phi$ formulation.

\section{Orthonormal Polynomial Projection Quantization-Bounding Method}

The preceding EMM analysis requires the use of nonlinear convex optimization for its implementation; otherwise referred to as semidefinite programming (SDP). Our available codes are fortran based, linear convex optimization adaptations of SDP, and limited by machine precision. Mathematica has recently implemented SDP algorithms. Application of these will improve the previous results by allowing us to work with arbitrary precision.

One limitation of EMM-$\Psi^2$ is that it is only applicable to one dimensional systems, because these are the only ones that admit a linear differential equation for the probability density, $\Psi^2(x)$. Such differential representations are required in order to generate the corresponding MER relation.

There is no known linear partial differential equation for the multidimensional probability density. In attempting to circumvent this difficulty, a new bounding method, purely algebraic in its implementation, was recently discovered.$^{17}$ We are able to apply it to arbitrary discrete states of multidimensional, bosonic or fermionic systems. It also can be extended to non-hermitian systems. We apply this procedure, referred to as the Orthonormal Polynomial Projection Quantization Bounding Method (OPPQ-BM) to the translated spiked harmonic oscillator, as given in Eq.(9). We will also generate the corresponding wavefunctions, as $b \rightarrow \infty$.

For the sake of completeness, we provide a comprehensive overview of the underlying OPPQ formalism. There are, essentially, two formulations of the OPPQ expansion. The first is as an eigenenergy approximation method. It is referred to as the Orthonormal Polynomial Projection Quantization - Approximation Method (OPPQ-AM).$^{18,19}$ The second, and preferred formulation, is the OPPQ-Bounding Method (OPPQ-BM). This formulation, OPPQ-BM, will produce tight bounds. Additionally, it also provides an alternate eigenenergy approximation ansatz that is more theoretically complete than OPPQ-AM, for reasons clarified below. We refer to the eigenenergy approximants, within OPPQ-BM, as {\it OPPQ-BM eigenenergy estimates}.

\subsection {The OPPQ Representation}

The OPPQ representation involves combining a particular type of basis expansion with the underlying MER relation for the system under consideration. In Eqs.(27,28) we developed a MER relation for the wavefunction configuration, ${\tilde \Psi}(\chi) \equiv \chi^{-2}\Psi(\chi)$.  We will develop the OPPQ analysis for this configuration space representation.

Consider the basis expansion
\begin{eqnarray}
{\tilde \Psi}(\chi) = \sum_{n=0}^\infty c_n {\cal B}_n(\chi)
\end{eqnarray}
where 
\begin{eqnarray}
 {\cal B}_n(\chi) = P_n(\chi) R(\chi),
\end{eqnarray}
for a positive reference function, $R(\chi) > 0$. The $P_n(x)$ are the orthonormal polynomials of a particular weight, $W(\chi)$. There are two options, $W(\chi) \in \{ R^2(\chi), R(\chi)\}$. We will refer to each as OPPQ-I and OPPQ-II, although our preferance is for the second, since there are more options in the selection of the weight.

If we choose the first option (OPPQ-I), and make the $P_n(\chi)$'s orthonormal relative to the weight $R^2$, then the basis $\{{\cal B}_n(\chi)\}$ is orthonormal, $\langle {\cal B}_m|{\cal B}_n\rangle = \delta_{m,n}$. For this case, we require that there be a MER relation for $\Phi(\chi) \equiv R(\chi) {\tilde \Psi}(\chi)$. If so, then we can generate the OPPQ projection coefficients exactly. The details of this will become evident when we consider the preferred formulation, OPPQ-II. Each formulation, OPPQ-I or OPPQ-II, has special advantages. We prefer OPPQ-II because, as alredy stated, it allows for greater flexibility in the choice of the weight, $W(\chi) = R(\chi)$. 

There are two immediate examples of  the superiority of the OPPQ-II formulation. First,  it allows for the use of piece-wise continuously differentiable weights. This becomes difficult within OPPQ-I, since the lack of differentiablity for the weight may complicate the generation of a MER relation within the $\Phi$ representation.

The other example of OPPQ-II's superiority is more profound. In reality, one does not need the explicit form for the weight; only its power moments are required (to high accuracy). This is because the orthonormal polynomials, $P_n$, are determined solely by the power moments of the weight. Thus, it is possible to solve for the power moments of the ground state (either through EMM or OPPQ-BM); and implement OPPQ-II, generating bounds to all the discrete state energies. The advantage in this is that the fastest convergence for OPPQ-AM or OPPQ-BM is achieved for weights very close to the asymptotic form of the physical states. Clearly this is impossible within the OPPQ-I formulation, since one needs the explicit form for the weight, in order to generate the required MER expression.

We proceed to develop the OPPQ-II representation. We assume (as is the case for the system in Eq.(9)) that the wavefunction configuration ${\tilde \Psi}(\chi)$ admits a MER representation (i.e. Eq.(28)). 

Let us take $P_n(\chi)$ to be the orthonormal polynomials for the weight $R(\chi)$. They are represented as

 \begin{eqnarray}
 P_n(\chi) = \sum_{j=0}^n \Xi_j^{(n)}\chi^j,
 \end{eqnarray}
and satisfy the orthonormal relation relative to the weight, $R(\chi)$:
\begin{eqnarray}
\langle P_{n_1}|R|P_{n_2}\rangle = \delta_{n_1,n_2}.
\end{eqnarray}
Based on this orthonormality condition, the configuration space basis ${\cal B}_n(\chi) \equiv P_n(\chi) R(\chi)$ in Eq.(55) is non-orthogonal, $\langle {\cal B}_{n_1}|{\cal B}_{n_2}\rangle \neq \delta_{n_1,n_2}.$

The OPPQ-II (henceforth simply referred to as OPPQ) projection coefficients can now be exactly calculated  through the underlying MER:
\begin{eqnarray}
c_n & = & \langle P_n|{\tilde\Psi}\rangle, \nonumber \\
& = & \sum_{j=0}^n\Xi_j^{(n)}
u(j),
\end{eqnarray}
where $u(p) = \int_0^\infty d\chi \ \chi^p {\tilde\Psi(\chi)}$ from Eq.(27). From Eq.(28) we can obtain the expression $u(p) = \sum_{\ell = 0}^{m_s} M_E(p,\ell) u_\ell$. Upon substituting we obtain
\begin{eqnarray}
c_n(E,{\overrightarrow u}) & = &
 \sum_{j=0}^n\Xi_j^{(n)}
\Big(\sum_{\ell = 0}^{m_s} M_E(j,\ell) \ u_\ell\Big) \nonumber \\
& = & \sum_{\ell = 0}^{m_s}\Lambda_{\ell}^{(n)}(E) u_\ell, \nonumber \\
& = & {\overrightarrow \Lambda}_E^{(n)}\cdot {\overrightarrow u},
\end{eqnarray}
where $\Lambda_{\ell}^{(n)}(E) = \sum_{j=0}^n \Xi_j^{(n)}M_E(j,\ell)$, and
${\overrightarrow u} = (u_0,u_1,\ldots, u_{m_s})$. We stress that the projection coefficients are known exactly as functions of the energy and missing moments.

\subsection{The OPPQ-Quantization Condition}
Consider the following integral expression for the physical configurations, assuming that the asymptotic properties of the adopted weight makes the  integral finite:

\begin{eqnarray}
I[{\tilde \Psi},R] \equiv \int_0^{\infty}d\chi \frac{{\tilde \Psi}^2}{R} = \sum_{n = 0}^\infty c_n^2(E, {\overrightarrow u}).
\end{eqnarray}
We also require that the asymptotic form for the weight also result in this integral becoming infinite for unphysical solutions to the Schrodinger equation. Accordingly, the OPPQ quantization condition becomes

\begin{eqnarray}
I[{\tilde \Psi},R] = \begin{cases}
finite \iff E = E_{phys} \ and {\overrightarrow u} = {\overrightarrow u}_{phys} \\
\infty \iff E \neq E_{phys}\ or \ {\overrightarrow u} \neq {\overrightarrow u}_{phys}.
\end{cases}
\end{eqnarray}
This corresponds, effectively, to a {\it shooting method} in the $E \times {\overrightarrow u}$ parameter space. Fortunately, the OPPQ formalism will reduce this to a minimization problem within the energy parameter space.

\subsection{Selection of the OPPQ Weight}

In order for Eq.(62) to hold, the weight must not decrease much faster than the asymptotic form of the physical solutions:

\begin{eqnarray}
\lim_{\chi \rightarrow \infty} \Big | \frac{{\tilde \Psi}(\chi)}{R(
\chi)}\Big | = C(\chi),
\end{eqnarray}
provided $\int_0^\infty d\chi \ |\Psi(\chi)| C(\chi) < \infty$, for all physical states, and infinite, otherwise. Given, this, the preferred choice of weight is one that mimics the asymptotic form of the physical solutions. 
\begin{eqnarray}
C(\chi) \sim {\tilde \Psi}_{phys}(\chi).
\end{eqnarray}
We have discovered that this leads to the fastest convergence.

Another important observation follows. If Eqs.(63,64) are adopted for the weight, then since one can choose the weight to satisfy $\frac{1}{R(\chi)} > 1$ and we also have $\lim_{\chi\rightarrow \infty}\frac{1}{R(\chi)} = \infty$, it follows that  $\langle {\tilde \Psi}|{\tilde \Psi}\rangle < {\cal I}[{\tilde \Psi},R]$. Thus, unphysical (i.e. non $L^2$) configurations have an infinite quantization integral (i.e. Eq.(61)). If the quantization integral is finite, then the associated configuration must be $L^2$; however, $R$ is chosen to make this always the case.

Once the positive (or nonnegative) weight is selected, $R \geq 0$, if it does not correspond to the classic orthogonal polynomials (i.e. Laguerre, Legendre, Hermite, etc.), then the orthonormal polynomials can be determiend through a Cholesky decomposition of the positive Hankel moment matrix of the weight. Thus, given the Hankel moment matrix for the weight, $W(i,j) = \int_0^\infty d\chi \ \chi^{i+j} R(\chi) \equiv {\cal C}{\cal C}^\dagger > 0$, we use Cholesky decomposition to generate:
\begin{eqnarray}
{\cal C}^\dagger {\overrightarrow \Xi}^{(n)} = {\hat {\cal E}}^{(n)},
\end{eqnarray}
where ${\hat {\cal E}}^{(n)}$ is the unit vector tuple. This then gives us the desired expression
$\langle {\overrightarrow \Xi}^{(m)}|{\cal C}{\cal C}^\dagger|{\overrightarrow \Xi}^{(n)}\rangle = \delta_{m,n}$.

We remind the reader that once the Hankel moment matrix for the weight is known, its orthonormal polynomials (including the multidimensional extension of this analysis) can be generated. Thus, if one can determine the power moments of the ground state wavefunction (for bosonic systems) through other method (i.e. EMM, or OPPQ-BM with a different weight), then one can implement OPPQ-BM with the ground state as a weight. This usually yields the fastest converging results within the OPPQ framework.

\subsection{The Normalization Condition on the Missing Moments}

Assume that the missing moments are normalized through some appropriate constraints:
\begin{eqnarray}
{\cal C}[{\overrightarrow u}] = 1.
\end{eqnarray}
For one space dimension problems, the natural normalization is imposing the unit vector constraint $\sum_{\ell = 0}^{m_s} u_\ell^2  = 1$. 

For multidimensional systems, with an infinite number of missing moments (although in a hierarchical manner), we require different types of normalizations. For the quadratic Zeeman problem$^{17}$ we used a very simple, linear normalization, corresponding to ${\cal C}[{\overrightarrow u}] = u_0$. Of course, one assumes that the choice of normalization does not filter out a desired physical state. 

Other choices are possible. Thus, for an $m_s =3$ problem, one can simply impose unit vector normalization on the first two missing moments:  ${\cal C}[{\overrightarrow u}] = u_0^2+u_1^2$ . Many other choices are also possible. The following discussion implicitly assumes that some missing moment normalization has been imposed.

\subsection {OPPQ-BM: Factoring Out the Missing Moment Variables}

Define the positive, partial sums defining the quantization integral in Eq.(61):
\begin{eqnarray}
S_N(E,{\overrightarrow u}) = \sum_{n=0}^N c_n^2(E;{\overrightarrow u}).
\end{eqnarray}
These partial sums trivially form a positive increasing sequence:
\begin{align}
0 < & S_{N}(E,{\overrightarrow u}) < S_{N+1}(E,{\overrightarrow u}) < 
S_{N+2}(E,{\overrightarrow u}) < \ldots \nonumber \\ 
& < \begin{cases} 
finite,  \iff \  E = E_{phys} \ {\rm and} \ {\overrightarrow u} =  
{\overrightarrow u}_{phys}; \\
\infty, \iff \  E \neq E_{phys} \ {\rm or} \ {\overrightarrow u} \neq 
{\overrightarrow u}_{phys}.
\end{cases}
\end{align}
From Eq.(60) we can rewrite these partial sums as
\begin{eqnarray}
S_N(E,{\overrightarrow u}) & = & \langle {\overrightarrow u}|{\cal P}^{(N)}(E)|{\overrightarrow u}\rangle, \\
{\cal P}^{(N)}(E) & \equiv & \sum_{n=0}^N {\overrightarrow \Lambda}_E^{(n)} {\overrightarrow \Lambda}_E^{(n)},
\end{eqnarray}
which is a sum of dyads. Once $N \geq m_s+1$, the ${\cal P}^{(N)}$  will be positive matrices.

It is clear from Eq.(61) that the integral expression is a function of the energy and missing moments: $I[\Psi, R] = I[E; {\overrightarrow u}]$. In the infinite expansion limit (i.e. sum the entire positive series in Eq.(61)), the expression $I[E; {\overrightarrow u}]$ is infinite everywhere except at the physical values. Using this  as a guide, Eq.(62) tells us that at the physical energy value, the physical missing moments are the ones that yield a global minimum for $I[E_{phys}; {\overrightarrow u}]$. Therefore, one should focus on the global minimum within the missing moment space for $S_N(E;{\overrightarrow u})$ in Eq.(69).

Let us define, to order $N$, the expression:
\begin{eqnarray}
{\cal L}_N(E) = Inf_{\overrightarrow u}\{S_N(E,{\overrightarrow u})|
{\cal C}({\overrightarrow u}) = 1 \},
\end{eqnarray}
involving some missing moment normalization, as discussed previously.

In the case of unit vector, missing moment normalization, we have:

\begin{eqnarray}
{\cal L}_N(E) = \lambda_N(E) \equiv {\rm Smallest\ Eigenvalue\ of } {\cal P}_N(E),
\end{eqnarray}
if ${\cal C}({\overrightarrow u}) \equiv |{\overrightarrow u}|^2 = 1$.

If the missing moment constraint is other than that of unit normalization, then ${\cal L}_N(E)$ involves constrained quadratic form minimization (CQFM). The details are straightforward and discussed elsewhere.$^{17}$

It is straightforward to argue that in general, regardless of the chosen normalization, the ${\cal L}_N(E)$ form a positive increasing sequence:
\begin{eqnarray}
{\cal L}_N(E) < 
{\cal L}_{N+1}(E) < \begin {cases}  finite, \ if\ E=E_{phys}, \\
\infty, \ if  \ E\neq E_{phys}.
\end{cases}
\end{eqnarray}
If a unit missing moment normalization is chosen, then
\begin{eqnarray}
\lambda_N(E) < \lambda_{N+1}(E) < \ldots.
\end{eqnarray}
In this case, as well as the general case in Eq.(73), these energy dependent functions form a nested, concaved upward sequence of functions. This is discussed in the numerical implementation of OPPQ-BM, as applied to Eq.(9).

Equations (73,74) define the  OPPQ-BM quantization conditions in the energy parameter space. Clearly the missing moment contribution has been factore out.

\subsection{The OPPQ-BM Eigenenergy Estimates}
We can argue that the local minima of the above expressions converge to the physical energy, $E_{phys}$. Define the local minima:
\begin{eqnarray}
\partial_E{\cal L}_N(E_N^{(min)}) = 0.
\end{eqnarray}
In a large neighborhood of the desired physical energy, $E_{phys}$ we can argue from the OPPQ quantization condition in Eq.(73), that the minimum values form an increasing positive sequence bounded from above:
\begin{eqnarray}
{\cal L}_N(E_{N}^{(min)}) < {\cal L}_{N+1}(E_{N+1}^{(min)}) < {\cal L}(E_{phys}) < \infty.
\end{eqnarray}
This follows from 
${\cal L}_N(E_{N}^{(min)}) < {\cal L}_N(E_{N+1}^{(min)}) < {\cal L}_{N+1}(E_{N+1}^{(min)})$. 

Thus, the local minima, within a large neighborhood of the corresponding physical state, must converge to it.
\begin{eqnarray}
lim_{N\rightarrow \infty} E_{N}^{(min)} = E_{phys}.
\end{eqnarray}
The approximants, $\{E_N^{(min)}\}$ will not, necessarily, converge monotonically; however, the function values ${\cal L}_N(E_{N})$ do converge monotonically. This then allows us to define a bounding procedure.

\subsection{The OPPQ Bounding Procedure} 

We have established that the local minima converge to the corresponding physical energy, $Lim_{N\rightarrow \infty}E_N^{(min)}= E_{phys}$, and the ${\cal L}(E_N^{(min)})$ form a  positive, increasing sequence, bounded from above. It then follows that any coarse upper bound to the latter sequence can be used to generate bounds for the  physical energies. 

Assume that a coarse upper bound has been empirically determined

\begin{eqnarray}
{\cal L}(E_N^{(min)}) < {\cal L}(E_{N+1}^{(min)}) < \ldots < {\cal B}_U.
\end{eqnarray}
Due to Eq.(73) one will always find energy roots to the equations:
\begin{eqnarray}
{\cal L}_N(E_N^{(L)}) = {\cal L}_N(E_N^{(U)}) = {\cal B}_U,
\end{eqnarray}
where $E_N^{(L)} < E_N^{(U)}$. These define the lower and upper bounds, respectively to the physical energy:
\begin{eqnarray}
E_N^{(L)} < E_{phys} < E_N^{(U)},
\end{eqnarray}
converging in the infinite limit
\begin{eqnarray}
Lim_{N\rightarrow \infty} (E_N^{(U)} - E_N^{(L)}) = 0^+.
\end{eqnarray}

It is important to stress that the Rayleigh - Ritz (RR) procedure, which yields converging upper bounds to the discrete state energies, has no criteria by which to empirically determine the accuracy of these upper bounds (despite the fact that they might be manifesting a convergent behavior). The OPPQ approach does. This is an important but subtle distinction between OPPQ-BM and RR. 

\section{The OPPQ-Approximation Method: A Quicker Eigenenergy Estimation Method}

Given that the positive series in Eq.(61) must converge for the physical parameter values, according to Eq.(62), it follows that

\begin{eqnarray}
Lim_{n\rightarrow \infty} c_n(E_{phys},{\overrightarrow u}_{phys}) = 0.
\end{eqnarray}

Since $c_n(E,{\overrightarrow u}) = {\overrightarrow \Lambda}_E^{(n)}\cdot {\overrightarrow u}$, we can develop an $(m_s+1)$ dimensional determinantal secular condition to approximate the physical energies: 

\begin{eqnarray}
{\overrightarrow \Lambda}_E^{(N-\ell)}\cdot {\overrightarrow u} = 0,
\end{eqnarray}
for $0 \leq \ell \leq m_s$ and $N \rightarrow \infty$. That is 
\begin{eqnarray}
Det\Big( \Lambda_{\ell_2}^{(N-\ell_1)}(E)  \Big) = 0, \cr
\end{eqnarray}
can be used to approximate the physical energies. This approach works very well, particularly if the weight, ``$R$'', mimics the asymptotic form of the physical states. We generally refer to this determinantal secular equation as the OPPQ-Approximation Method (OPPQ-AM), although the bounding procedure also introduces its own approximation method, through the local minima in Eq. (75). 

One immediate disadvantage of OPPQ-AM is that there is no guarantee that real energy roots will be generated. That is, the energy roots can have small imaginary parts that vanish in the asymptotic limit, $N \rightarrow \infty$. However, the OPPQ-BM Energy Estimates in Eq.(75) will always be real, for hermitian systems.

There is no guarantee that Eq.(82) will always generate the physical approximants. Spurious (nonconvergent) energies may result. This follows from the trivial result that the convergence of a positive series $\sum_{n=0}^\infty a_n$ has as a necessary condition that $\lim_{n\rightarrow \infty}a_n = 0$; however, this is insufficient. 

Despite the above, the physical energies and missing moments must satisfy Eq.(82), and exhibit a convergent behavior with the expansion order. This is a consequence of OPPQ-BM; which can also be a good validator for the OPPQ-AM results.

\section{OPPQ-Numerical Results}

The adopted strategy in this work is to use Eq.(84), and to confirm the results through the OPPQ-BM bounds. Generally, the OPPQ-AM approximants appear to converge faster to the physical energy than the approximants generated through the OPPQ-BM formalism (i.e. Eq.(75)), as well as the generated bounds.

Given that $\Psi_{phys}(\chi) \sim \chi^{\frac{3} {2}}A_1(\chi) $, as argued in Eq.(19), and ${\tilde \Psi}(\chi) = \chi^{-2}\Psi(\chi)$, as argued in Eqs.(27-28), the desired OPPQ weight (within the $\tilde \Psi$ representation) is chosen to be

\begin{eqnarray}
R(\chi) = \chi^{-{1/2}}exp\big(-\frac{1}{2}(\chi-b)^2 \big).
\end{eqnarray}

To generate the orthonormal polynomials, we will use a Cholesky decomposition of the Hankel  matrix $\omega(i+j)$ constructed from the power moments of the weight $\omega(p) = \int_0^\infty d\chi \ \chi^p R(\chi)$. 

One might be tempted to use the integral identity
\begin{equation}
\int_0^\infty d\chi \chi^p R(\chi) = \Gamma(p+1/2) e^{-b^2/4}  D_{-(p+1/2)} (-b),
\end{equation}
involving the ParabolicCylinderD function, $D_\nu(z)$, to compute the power moments of the weight, in order to generate the orthonormal polynomials. However, an alternate, and more efficient, strategy is to use the differential equation for the weight

\begin{equation}
\chi R'(\chi) = -\frac{1}{2} R(\chi) -(\chi^2-b\chi)R(\chi),
\end{equation}
in order to generate a recursive, MER, for the corresponding power moments:

\begin{equation}
\omega(p+2) = (p+\frac{1}{2}) \omega(p) + b \omega(p+1),
\end{equation}
$p \geq 0$. High precision calculations for $\{\omega(0),\omega(1)\}$ generate all the other power moments.
\subsection{OPPQ-Approximation Method}
\subsubsection{Recovering the Exact Energies for $b = 0$}

Generally, the OPPQ-AM secular condition in Eq.(84), $Det\Big( \Lambda_{\ell_2}^{(N-\ell_1)}(E)\Big) = 0$, will reveal the energies for an exactly solvable system, as is the case for $b = 0$. The structure of the determinant, as given in Table 4 confirms this.

To explain the results of Table 4, we recall that from the analysis leading to Eq.(38), we established that the physical solutions (for $b = 0$) are of the form in Eq.(44): $\Psi_n(\chi) = P_n(\chi^2)\chi^{\frac{3}{2}} exp(-\chi^2/2)$, involving the orthogonal polynomials $P_N(\xi)$ relative to the weight $\xi exp(-\xi)$. 

The OPPQ-AM ansatz generates the orthonormal polynomials of $R(\chi) = \chi^{-\frac{1}{2}} exp(-\chi^2/2)$. Let us refer to these as $Q_j(\chi)$, for the purposes of this discussion. We emphasize that these are polynomials in $\chi$ not $\chi^2$. Clearly, for each $\Psi_n$ eigenstate, we can express $\chi^2P_n(\chi^2) = \sum_{j=0}^{2n+2} d_{n;j}Q_j(\chi)$, or
$\chi^{\frac{3}{2}}P_n(\chi^2) = \chi^{-\frac{1}{2}}\sum_{j=0}^{2n+2} d_{n;j}Q_j(\chi)$. Therefore, when we implement OPPQ-AM, in Eq.(84), this should get the exact energies when $N \geq 2n+2+m_s+1$. 

Since $m_s = 3$ for Eq.(28), when $b \neq 0$ (and effectively an $m_s = 1$ when $b = 0$) it follows that $N-4 \geq 2n$, will give the exact energies within OPPQ-AM. We see in Table 4 that the $n = 0$ state is exactly determined for $N \geq 4$. The first excited state (i.e. $E_1 = 4$) is exactly generated for $N \geq 6$, etc. 
\begin{table}
\caption{{}
$OPPQ-AM: Det\Big( \Lambda_{\ell_2}^{(N-\ell_1)}(E) \Big)$, for $b = 0$.
}
\begin{ruledtabular}
\begin{tabular}{rl}
\hline
$N$ & $Det^{(N)}(E) \propto$ \\
\hline
4 & $(E-2) $\\
5& $  (E-2)(E-4.25{\overline 4}) $   \\
6&   $(E-2)(E-4)(E-7.46{\overline 3}) $   \\
7&   $(E-2)(E-4)(E-5.708)(E-12.632) $   \\
8&   $\Pi_{n=1}^3(E-2n)\times Poly(degree\ 2)$\\  
10& $\Pi_{n=1}^4(E-2n)\times Poly(degree\ 3)$\\  
20& $\Pi_{n=1}^9(E-2n)\times Poly(degree\ 8)$\\
30&$\Pi_{n=1}^{14}(E-2n)\times Poly(degree\ 13)$\\
\hline
\hline
\end{tabular}
\end{ruledtabular}
\end{table}

\subsubsection{Numerical OPPQ-AM Results}

In Table 5 we show the results of implementing OPPQ-AM for various values of $b: 0 \rightarrow 10$. The results in Table 5 are stable (i.e. convergng) to many more digits than given; and correspond to using moment expansion order $P_{max} = 100$.  We note the exact accuracy of the energies for the $b = 0$ case. Comparing the $E_0$ entries for $b = .5,1, 5, 10$ with those in Table 2 (derived through EMM), shows that OPPQ-AM results are consistent with the tight EMM bounds. The same applies for the excited states, with bounds reported in Table 3, also derived through EMM-$\Psi^2$. The results in Table 5 are graphically illustrated in Fig. 1.

\begin{table}
\caption{
$OPPQ-AM, P_{max} = 100$.
}
\centerline{
\begin{tabular}{rrrrr}
\hline
$b$ & $E_0$    &  $E_1$  &   $E_2$  & $E_3$\\
\hline
0& 2 & 4  & 6 & 8 \\
.5&  1.4292927197    &    3.184017114	& 4.987971463 &	6.820440707\\
1.0&1.0331033239	    & 2.557261915	&4.169923329	&  5.837014390\\
1.5&0.7847675572   &2.107433725	&3.538491138	&5.044354682\\
2.0&0.6481322228  &	1.816590914	 &3.084658976	&4.436894490\\
2.5&0.5818553905	&1.655297046	&2.794166923	&4.007820744\\
3.0&0.5509509520	   &1.580121756	&2.638483895	&3.743614149\\
3.5&0.5351717068	&1.547741639	&2.570043163	&3.611459829\\
4.0&0.5259688826	&1.532318972	&2.541876785	&3.557419442\\
4.5&0.5200471427	&1.523670892	&2.528557218	&3.535449712\\
5.0&0.5159807819	&1.518222436	&2.521046746	&3.524694536\\
5.5&0.5130560157	&1.514525084	&2.516293284	&3.518454199\\
6.0&0.5108767399	&1.511882954	&2.513054951	&3.514433855\\
6.5&0.5092067728	&1.509920644	&2.510731970	&3.511660509\\
7.0&0.5078974472	&1.508418730	&2.509000073	&3.509651653\\
7.5&0.5068510767	&1.507241028	&2.507669464	&3.508141930\\
8.0&0.5060011798	&1.506298942	&2.506622186	&3.506974079\\
8.5&0.5053011643	&1.505532597	&2.505781384	&3.506049402\\
9.0&0.5047175513	&1.504900234	&2.505095024	&3.505303077\\
9.5&0.5042257611	&1.504371940	&2.504526749	&3.504690917\\
10.0&0.5038074053	&1.503925798	&2.504050459	&3.504181861\\
 \hline
\hline
\end{tabular}}
\end{table}

\begin{figure*}
\includegraphics{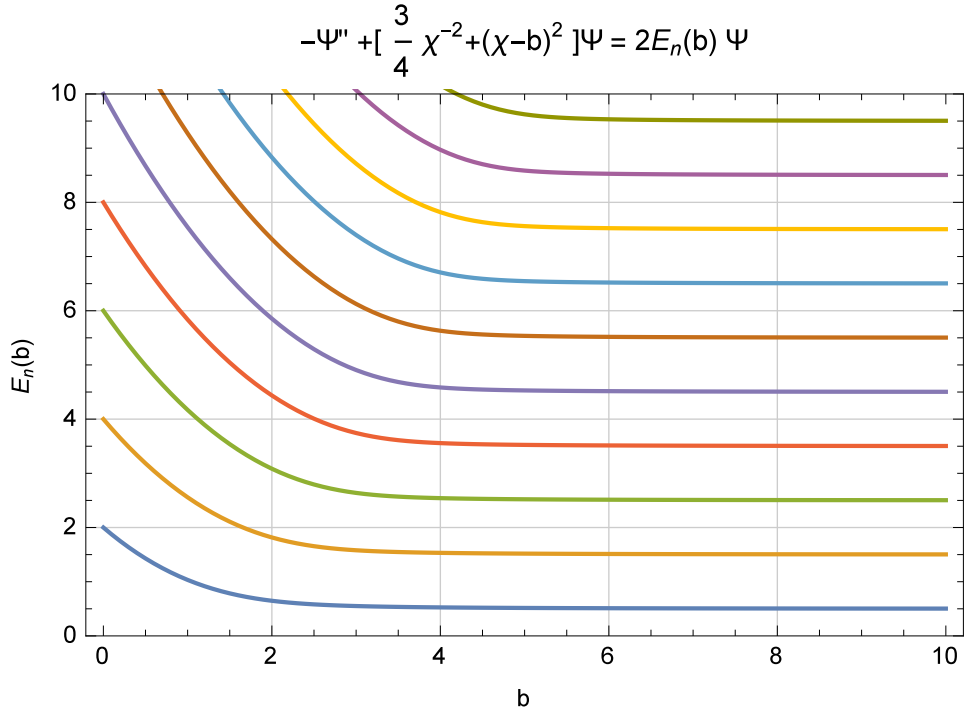}
\caption{\label{fig.wide1}OPPQ-AM Results from Table 5: $E_n(b)$, for $0 \leq n \leq 9$ .}
\end{figure*}

\subsection{Results for the OPPQ-Bounding Method (BM)}

\subsubsection{The Nested $\lambda_n(E)$ Function Sequence}

    In Figs. 2-5, we show the ${\cal L}_n(E) \equiv \lambda_n(E)$ nested function sequence in Eq.(73), for $b=.5$ and the first four energy levels given in Table 5.  It is important to recognize that the extrema locations in the energy parameter variable (i.e. $\partial_E\lambda_n(E_n^{min}) = 0$), do not necessarily behave in a monotonic manner until (perhaps) they get close to the physical  energy value. However, the value of $\lambda_n(E_n^{(min)})$ do converge, from below, monotonically to the correct physical value, in the infinite expansion limit. This is best appreciated from Table 6, which captures all the extrema information in Figures 2-5 for the case $b = 0.5$.

\begin{figure}
\center{\includegraphics{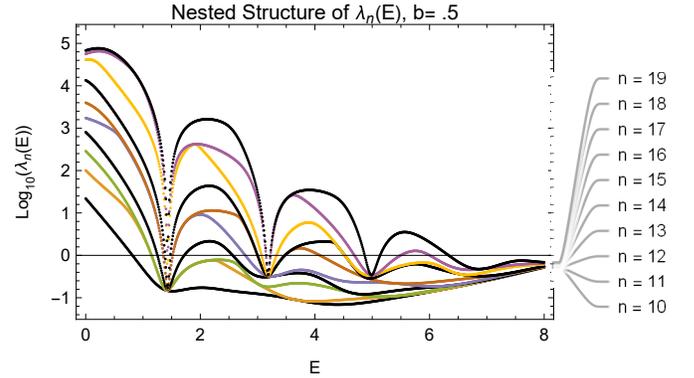}}
\centering{\caption{Nested $\lambda_n(E)$ sequence, $n = 10, 11, \ldots, 19$, $0 \leq E \leq 8$, and $b = .5$ (i.e. Eq.(73))}}
\end{figure}

\begin{figure}
\center{\includegraphics{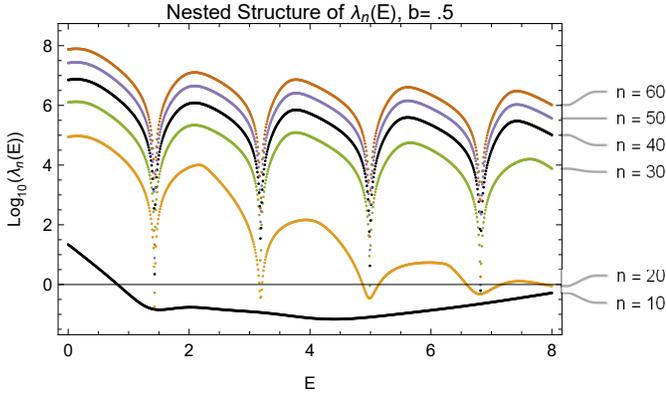}}
\centering{\caption{Nested $\lambda_n(E)$ sequence, $n = 10, 20, \ldots, 60$, $0 \leq E \leq 8$, and $b = .5$ (i.e. Eq.(73))}}
\end{figure}

\begin{figure}
\center{\includegraphics{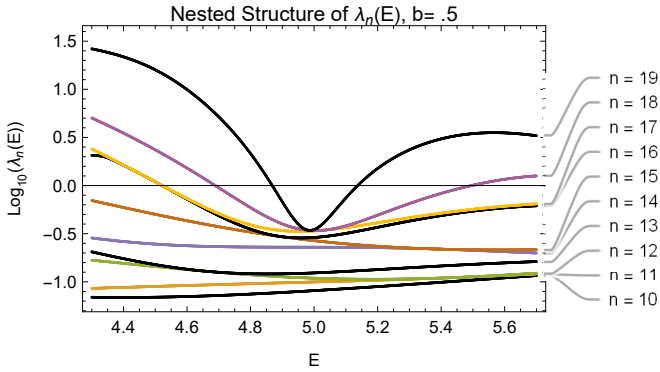}}
\centering{\caption{Nested $\lambda_n(E)$ sequence, $n = 10, 11, \ldots, 19$, $4.4 \leq E \leq 5.6$, and $b = .5$ (i.e. Eq.(73))}}
\end{figure}

\begin{figure}
\center{\includegraphics{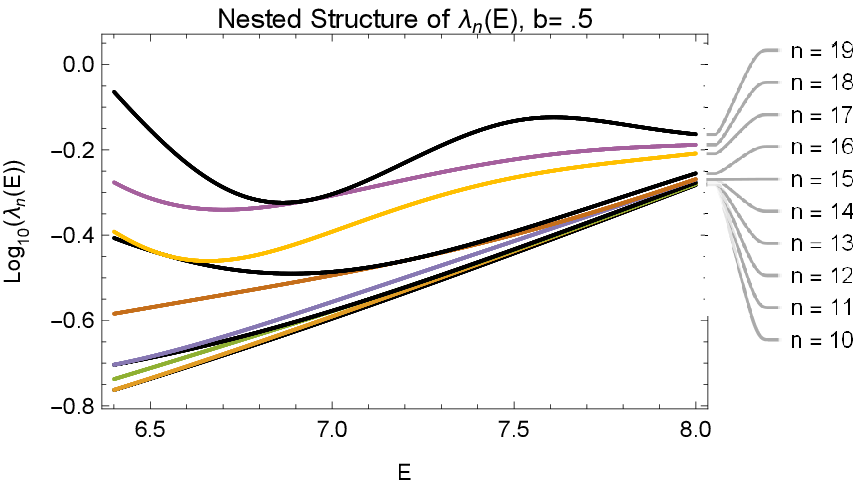}}
\centering{\caption{Nested $\lambda_n(E)$ sequence, $n = 10, 11, \ldots, 19$, $6.5 \leq E \leq 8.0$, and $b = .5$ (i.e. Eq.(73))}}
\end{figure}

\begin{table*}
\caption{\label{}  OPPQ-BM Energy Approximants: $\partial_E\lambda_N(E_{N,n_q}) = 0$,  and $\lambda_{log} \equiv Log_{10}\big(\lambda_N(E_{N;n_q})\big)$, for $ b = .5$}
\begin{ruledtabular}
\begin{tabular}{rllllllll}
\hline
N& $ E_{N;0} $& $\lambda_{log}(E_{N;0}) $& $E_{N;1}$ &$ \lambda_{log}(E_{N;1}) $ & $E_{N;2} $& $\lambda_{log}(E_{N;2})$& $E_{N;3}$ &$ \lambda_{log}(E_{N;3}) $\\ \hline

10&  1.5150470& -0.84559280  & 4.3969969 & -1.1623635        &     &    &      &        \\
11&  1.4199646 & -0.82806681&  3.9889962  &  -1.0852239       &   &       &    &     \\
12 & 1.4156228 & -0.80336151&  3.1875626 &  -0.74234366 & 5.1889348 & -0.97331160&  & \\
13 & 1.4301144  & -0.79832277&  3.1214643 & -0.51419926 & 4.8564978 & -0.91784021&  &\\
14 & 1.4290630 &  -0.79825893& 3.2393572 & -0.49476172 &6.0745714 &   -0.73221637&  & \\
15 & 1.4289479 &  -0.79757698& 3.1808773 & -0.48495873 & 5.5858837 &  -0.66854618&  & \\
16 & 1.4294188&  -0.79745638&  3.1790069 &-0.47258623 & 4.9600850 &  -0.54416892 & 6.8883817  & -0.49025962\\
17 & 1.4293205 &  -0.79744519& 3.1863572 &  -0.47115027&4.9543391 &-0.47566983&  6.6591851 & -0.46057027\\
18 & 1.4292787 &  -0.79740046&  3.1841563  & -0.47087295 & 5.0040638 & -0.46768132&6.7008468 & -0.34041178\\
19 & 1.4292932 & -0.79738417&   3.1837027&  -0.47025848&4.9868242  & -0.46552191& 6.8641735 & -0.32407160\\
20 & 1.4292967&  -0.79738248 &  3.1840333 & -0.47013608 & 4.9851930  & -0.46121438&  6.8069551 &-0.31888442\\
\hline
30&  1.4292931 &-0.79738040& 3.1840182 &-0.47012365& 4.9879738&-0.46026542&6.8204428&-0.30655075\\
40& 1.4292928& -0.79738016&3.1840173 &-0.47012319 &4.9879718&-0.46026461& 6.8204411 &-0.30655027\\
50& 1.4292927 & -0.79738012&3.1840172 &  -0.47012312&4.9879716&-0.46026449&6.8204408&-0.30655019\\
60& 1.4292927&-0.79738011&3.1840171&-0.47012310&4.9879715 &-0.46026446&6.8204408&-0.30655017\\
70 & 1.42929272246 &-0.7973801116&  3.18401712169&  -0.4701230941& 4.98797147862& -0.4602644456&6.82044072541 & -0.3065501612\\
80 & 1.42929272103 &-0.7973801106&  3.18401711764&  -0.4701230920& 4.98797147036& -0.4602644418&6.82044071591 & -0.3065501588\\
90 & 1.42929272042 & -0.7973801101&3.18401711589& -0.4701230911&    4.98797146679& -0.4602644402&6.82044071179& -0.3065501577\\
100&1.42929272012&  -0.7973801099&3.18401711506 &-0.4701230907&    4.98797146508& -0.4602644394&6.82044070983& -0.3065501572\\
150&1.42929271979&  -0.7973801096&3.18401711412 &-0.4701230902&    4.98797146316& -0.4602644386&6.82044070761&-0.3065501566\\
200&1.42929271976&  -0.7973801096&3.18401711403 &-0.4701230901&    4.98797146298& -0.4602644385&6.82044070740&-0.3065501566\\
250&1.429292719754&  -0.7973801096&3.18401711401 &-0.4701230901&    4.98797146294& -0.4602644385&6.82044070736&-0.3065501565\\
300&1.429292719752&  -0.7973801096&3.18401711401 &-0.4701230901&    4.98797146293& -0.4602644385&6.82044070735&-0.3065501566\\
350&1.4292927197517&  -0.7973801096&3.18401711400 &-0.4701230901&    4.98797146293& -0.4602644385&6.82044070734&-0.3065501566\\
\hline
${\cal B}_U$ &    & $10^{-.79738}$ &  & $10^{-.470123}$ &  &$10^{ -.460264}$&  &$10^{-.30655}$\\
\hline
\end{tabular}
\end{ruledtabular}
\end{table*}

\begin{table*}
\caption{\label{}  OPPQ-BM Upper and Lower Bounds: $E_N^{(L)} < E_{phys} < E_N^{(U)}$,  for $ b = .5$
}
\begin{ruledtabular}
\begin{tabular}{rllllllll}
\hline
$N$& $E_{0}^{(L)}$ & $E_{0}^{(U)}$ &$E_{1}^{(L)}$ & $E_{1}^{(U)}$  & $E_{2}^{(L)}$ & $E_{2}^{(U)}$&$E_{3}^{(L)}$ & $E_{3}^{(U)}$ \\
\hline
10&   1.355213912&     1.767314750  &  &        &     &    &      &        \\
50&   1.429292680  &   1.429292800&     3.184017055   &  3.184017276      &   4.987971200 &     4.987972000     & 6.820440400   & 6.820441200  \\
100& 1.4292927126 &  1.4292927276&   3.184017100   &  3.184017130 &        4.987971416  &    4.987971512&      6.820440664  &  6.820440752 \\
150& 1.4292927172 &  1.4292927224&   3.1840171096 &  3.1840171186 &       4.9879714476 &   4.9879714784&   6.8204406948  & 6.8204407212\\
\hline
 \hline
${\cal B}_U$ &    & $10^{-.79738}$ &  & $10^{-.470123} $&  &$10^{ -.460264}$&  &$10^{-.30655}$\\
\hline
\end{tabular}
\end{ruledtabular}
\end{table*}

\subsubsection{OPPQ-BM Energy Estimates}
The  local minima
\begin{eqnarray}
\partial_E\lambda_I(E_I^{min}) = 0,
\end{eqnarray}
define the $I$-th order (OPPQ-BM) approximant to the discrete state energy. These extrema are necessarily real. Also, one does not have to numerically determine these derivatives. It is straightforward to generate an algebraic procedure for generating the function $\partial_E\lambda_I(E)$, and then determine its zeroes, and in particular the local minima.$^{17}$ The results of this analysis are given in Table 6.

More specifically, the results in Table 5, based on OPPQ-AM, were generated based on a maximum moment expansion order of $P_{max} = 100$. With regards to the ground state energy, particularly at $b = .5$, we see that the OPPQ-AM results in Table 5 ($E_{gr}(b=.5) = 1.4292927197$)  concur with the EMM bounds in Table 2 ( $1.4292927197475 < E_{gr} < 1.4292927197522$) generated on the basis of maximum moment expansion order $P_{max} = 24$ (based on Eqs.[47-50] with $\sigma = 3$). However, within the OPPQ-BM's Approximation Ansatz (i.e. Eq. (73)), it takes $P_{max} = 150$ (i.e. Table 6) to achieve comparable results to OPPQ-AM for $P_{max} = 100$. It takes OPPQ-BM's Approximation Ansatz a $P_{max} = 350$ to compete with the EMM bounds in Table 2 (obtained with $P_{max} = 24$). Despite all this, EMM-$\Psi^2$ cannot give the same tightness of bounds for the excited states, as amply demonstrated in Table 3. We require OPPQ-BM to achieve decent bounds, surpassing those of EMM for the excited states. This is discussed below.
\\
\\

\subsubsection {\bf OPPQ-BM: Generating Eigenenergy Bounds}

The local minima in Eq.(89) also serve to define an increasing positive sequence that is bounded from above by the true physical energy counterpart:

\begin{eqnarray}
\lambda_I(E_I^{(min}) < \lambda_{I+1}(E_{I+1}^{(min})  < \ldots < \lambda_\infty(E_{phys}) < \infty.
\end{eqnarray}

One can use Eq.(90) to generate empirically converging bounds on the true physical energy. Thus, let ${\cal B}_U$ be any coarse upper bound to the  convergent expression (i.e. for a chosen discrete state):
\begin{eqnarray}
\lambda_I(E_I^{(min}) < \lambda_{I+1}(E_{I+1}^{(min})  < \ldots < \lambda_\infty(E_{phys}) < {\cal B}_U.
\end{eqnarray}
From Eq.(73),  it is straightforward to argue that there will always be $E$ values satisfying:
\begin{eqnarray}
\lambda_I(E_I^{(L)}) = \lambda_I(E_I^{(U}) = {\cal B}_U.
\end{eqnarray}
These will then correspond to converging lower and upper bounds to the  desired physical energy:
\begin{eqnarray}
E_I^{(L)} < E_{phys} < E_I^{(U)},
\end{eqnarray}
with
\begin{eqnarray}
Lim_{I \rightarrow \infty} \Big( E_I^{(U)}-E_I^{(L)} \Big) = 0^+.
\end{eqnarray}

The results of this bounding analysis on the  first four discrete states for $b = .5$, utilizing the coarse upper bounds ${\cal B}_U$ appearing at the bottom of Table 6, are given in Table 7. We note that these are far superior to those in Table 3, based on EMM-$\Psi^2$. Keeping these coarse ${\cal B}_U$ upper bounds, we can generate converging lower and upper bounds, as shown in Table 7.

\section {Wavefunction Reconstruction}

The OPPQ expansion of the wavefunction $\Psi(\chi) \equiv \chi^2 {\tilde \Psi}(\chi)$ is obtained from Eq.(55). The wavefunction reconstruction uses all the previously generated results (i.e. physical energy approximants, etc) but only uses the OPPQ-expansion coefficients, $\{c_n| n \leq 40\}$. We plot the results for the first four discrete state energy levels corresponding to $b=\{0,1,2,3,4,5\}$ in order to visualize the evolution of the wavefunctions relative to the anticipated harmonic oscillator solutions as $b \rightarrow \infty$. All of this is depicted in Figures 6-11. It will be noticed that relative to the point $\chi=b$, the wavefunctions become more symmetric, or antisymmetric, as $b \rightarrow 5$.

\begin{figure*}
\includegraphics{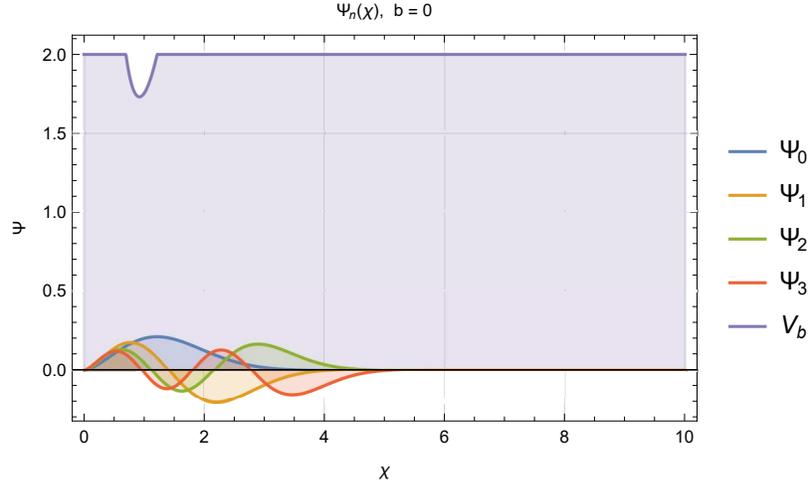}
\caption{ \label{fig:widew0}Discrete state   wavefunctions, $\Psi_n(\chi)$, for $n \in \{0,1,2,3\}$; and potential, $V_b(\chi) = \frac{1}{2}\Big(\frac{3}{4}\chi^{-2} + (\chi-b)^2\Big)$, for $b = 0$}
\end{figure*}

\begin{figure*}
\includegraphics{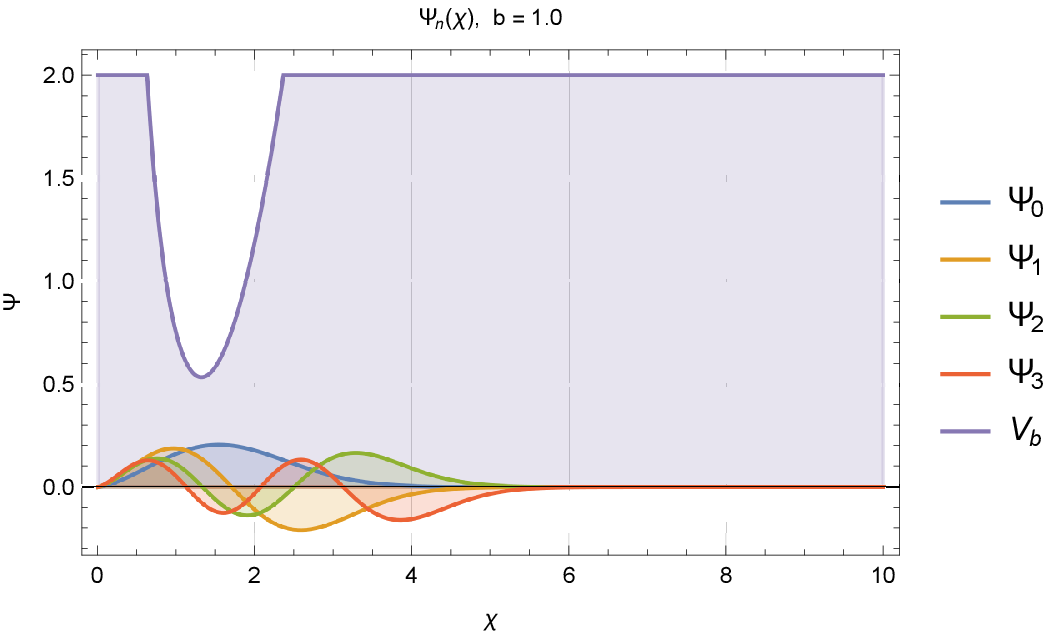}
\caption{ \label{fig:widew1} Discrete state wavefunctions, $\Psi_n(\chi)$, for $n \in \{0,1,2,3\}$; and potential, $V_b(\chi) = \frac{1}{2}\Big(\frac{3}{4}\chi^{-2} + (\chi-b)^2\Big)$, for $b = 1$}
\end{figure*}

\begin{figure*}
\includegraphics{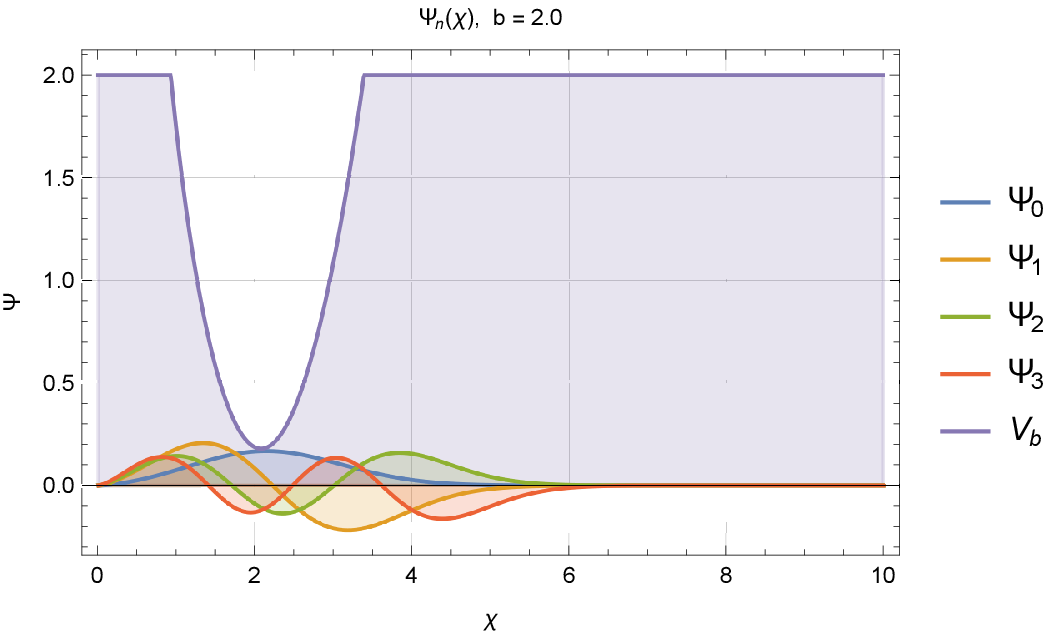}
\caption{ \label{fig:widew2} Discrete state wavefunctions, $\Psi_n(\chi)$, for $n \in \{0,1,2,3\}$; and potential, $V_b(\chi) = \frac{1}{2}\Big(\frac{3}{4}\chi^{-2} + (\chi-b)^2\Big)$, for $b = 2$}
\end{figure*}
\begin{figure*}
\includegraphics{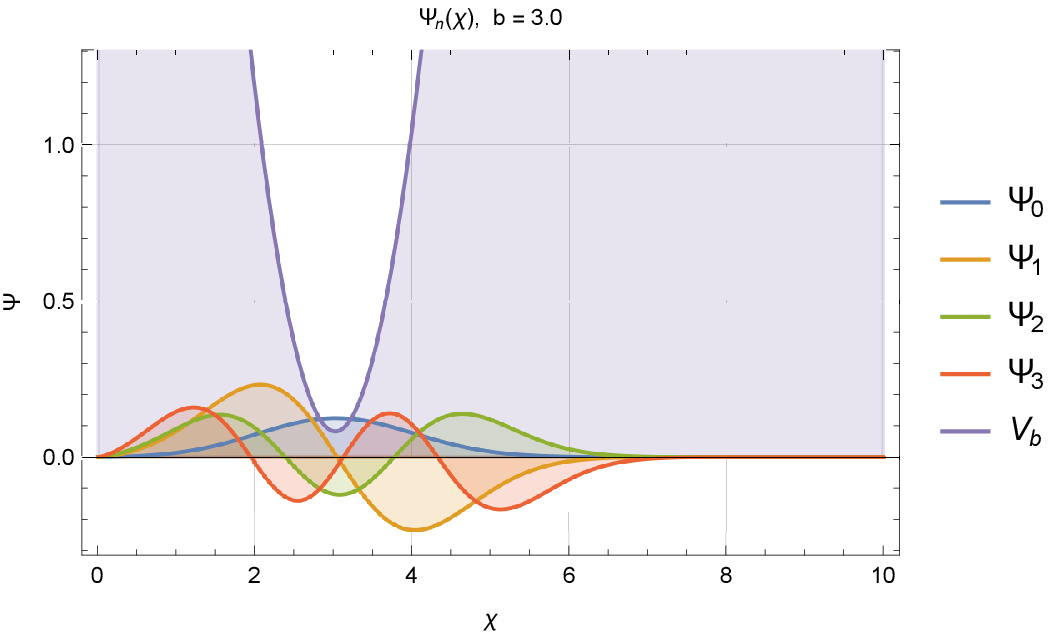}
\caption{ \label{fig:widew3} Discrete state wavefunctions, $\Psi_n(\chi)$, for $n \in \{0,1,2,3\}$; and potential, $V_b(\chi) = \frac{1}{2}\Big(\frac{3}{4}\chi^{-2} + (\chi-b)^2\Big)$, for $b = 3$}
\end{figure*}

\begin{figure*}
\includegraphics{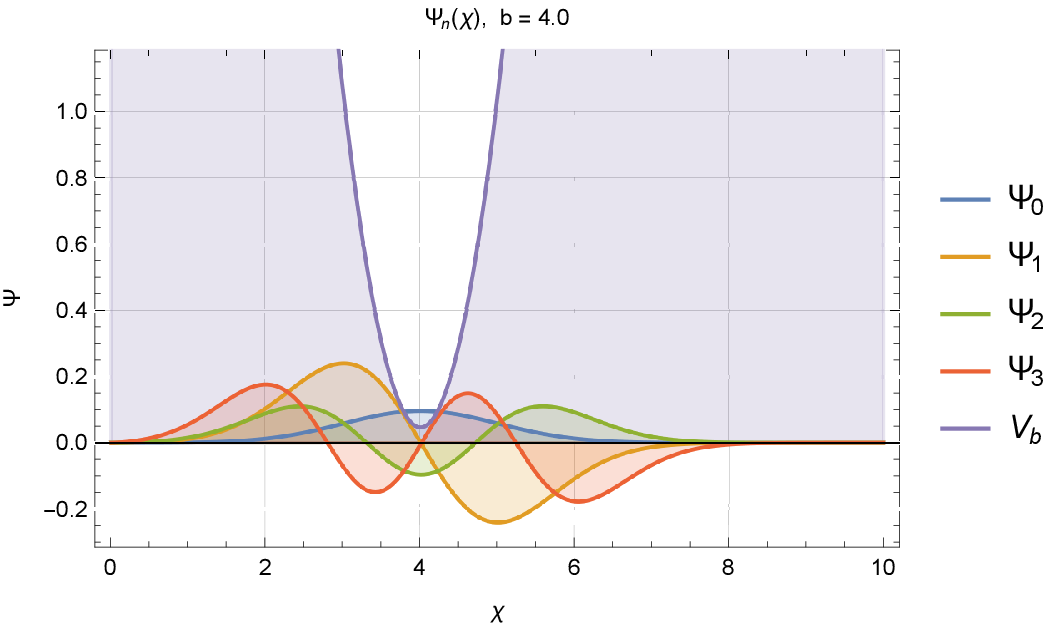}
\caption{ \label{fig:widew4} Discrete state wavefunctions, $\Psi_n(\chi)$, for $n \in \{0,1,2,3\}$; and potential, $V_b(\chi) = \frac{1}{2}\Big(\frac{3}{4}\chi^{-2} + (\chi-b)^2\Big)$, for $b = 4$}
\end{figure*}

\begin{figure*}
\includegraphics{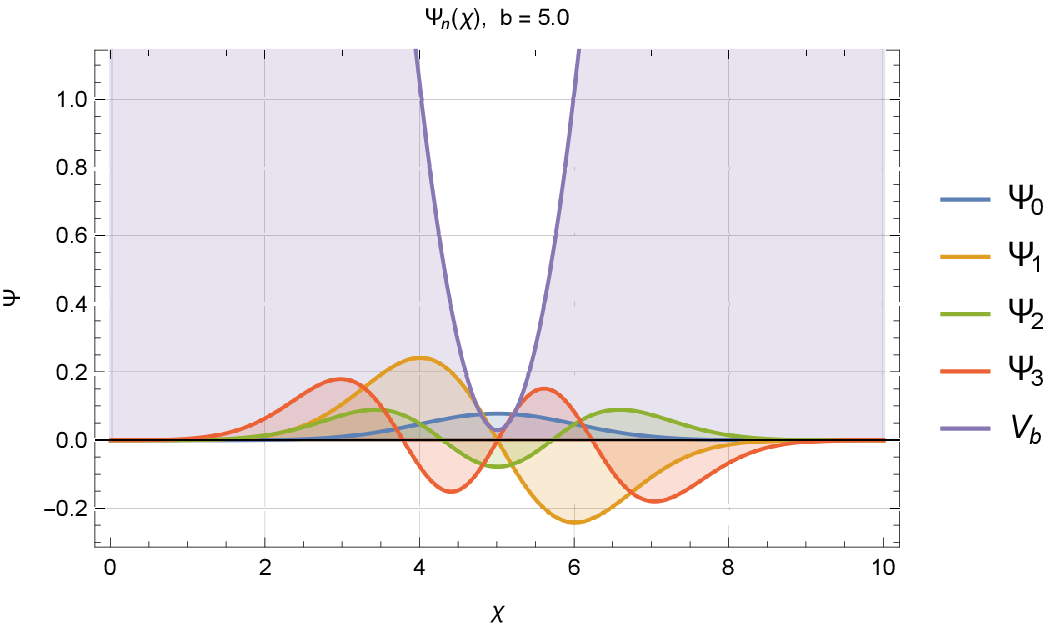}
\caption{ \label{fig:widew5} Discrete state wavefunctions, $\Psi_n(\chi)$, for $n \in \{0,1,2,3\}$; and potential, $V_b(\chi) = \frac{1}{2}\Big(\frac{3}{4}\chi^{-2} + (\chi-b)^2\Big)$, for $b = 5$}
\end{figure*}

\section{Conclusion}
We have examined, in numerical detail, the low lying discrete state energies of the translated {\it spiked harmonic oscillator}, generated from an {\it affine quantization} analysis of the {\it half harmonic oscillator} problem. Our methods are highly accurate, due to their abilities to generate tight bounds to the desired physical states. Our numerical results confirm that as the {\it wall slides away}, the system becomes that of the full harmonic oscillator problem on the real domain.
\section{Acknowledgement}

The author is deeply grateful to Dr. John R. Klauder for inspiring this work, and making valuable suggestions. 

\section{Appendix}
\subsection{Eigenstates of the Half-Harmonic Oscillator}

In this section we return to the {\it half harmonic oscillator} (HHO)

\begin{eqnarray}
-\partial_x^2 \Psi(x) + x^2 \Psi(x) = 2E \Psi(x),
\end{eqnarray}
for $x \geq 0$, and show how the {\it even states}  are dense with respect to the {\it odd states}. Although this is an anticipated result, the derivation is important in affirming the challenges of canonical quantization as applied to systems on bounded Cartesian domains.

Consider the subspaces
\begin{eqnarray}
S_{\cal E} & \equiv & \{ \Psi(x)| \Psi \in L^2(\Re^+,dx), \partial_x\Psi(0) = 0\},\\
S_{\cal O} & \equiv & \{ \Psi(x)| \Psi \in L^2(\Re^+,dx), \Psi(0) = 0\}.
\end{eqnarray}
  The kinetic energy operator, $-\partial_x^2$, is Hermitian within each, although the linear momentum operator, $-i\partial$, is only Hermitian within the $S_{\cal O}$ subspace. As noted earlier, the focus of {\it affine quantizaiton} is to identify the appropriate Hamiltonian that is Hermitian on the entire space, $L^2(\Re^+,dx)$.
 
 The  HHO problem admits $L^2$, normalized, eigenfunctions within these subspaces. They correspond to

\begin{eqnarray}
\Psi_n(x) = \begin{cases} {\cal E}_\eta (x) \equiv \frac{1}{\sqrt{2}}H_{n}(x)  e^{-\frac{x^2}{2}},\ n = 2\eta ;\\
{\cal O}_\eta (x) \equiv \frac{1}{\sqrt{2}}H_n(x)e^{-\frac{x^2}{2}}  ,\ n = 2\eta+1  \end{cases}
\cr
\end{eqnarray}
for $\eta = 0,1,\ldots$. The
  $\{H_n(x)\}$ are the Hermite polynomials relative to the weight $W(x) = exp(-x^2)$, satisfying the adopted normalization condition (on the entire real axis, $\Re$):
\begin{eqnarray}
\langle H_{n_1}|W|H_{n_2}\rangle_{\Re} = 2\delta_{n_1,n_2},
\end{eqnarray}
$n_{1,2} \geq 0$. 
 
  The corresponding eigenenergies are given by the well known expression, $E_n = (n+\frac{1}{2})$, for nonnegative integers, $n \geq 0$.

 \subsubsection{Completeness Properties on the Positive Real Axis}
 
   Restricted to the nonnegative real axis, the even and odd eigenfunctions in Eq.(98) are orthonormal solely amongst themselves (i.e. $ H_{2\eta}(x) exp(-\frac{x^2}{2}) = \sqrt{2}{\cal E}_\eta(x)$ and $\ H_{2\eta+1}(x) exp(-\frac{x^2}{2}) = \sqrt{2}{\cal O}_\eta(x) $):
\begin{eqnarray}
\langle H_{2\eta_1}|W|H_{2\eta_2}\rangle_{Re^{0,+}} & = & \delta_{\eta_1,\eta_2}, \\
\langle H_{2\eta_1+1}|W|H_{2\eta_2+1}\rangle_{Re^{0,+}} & = & \delta_{\eta_1,\eta_2}, \\
\langle H_{2\eta_1}|W|H_{2\eta_2+1}\rangle_{Re^{0,+}} &  \neq & 0.
\end{eqnarray}

The two sets of functions $\{{\cal O}_\eta(x)\}\cup \{{\cal E}_\eta(x)\}$ form a complete, orthonormal, basis for all $L^2$  functions on the real axis.
However, if restricted to the nonnegative real axis, the corresponding expressions are over-complete, for all $L^2$ functions on $\Re^{0,+}$:

\begin{eqnarray}
\{{\cal O}_\eta(x),{\cal E}_\eta(x)\} = \begin{cases} Complete \ on \ L^2(\Re,dx) \\
Over-complete \ on \ L^2(\Re^{0,+},dx). \\
\end{cases}
\end{eqnarray}

Let $\Psi_{phys} \in S_{\cal O}$. Then, in terms of the eigenfunctions of this subspace, we have
\begin{eqnarray}
\Psi_{phys}(x) = \sqrt{2}\sum_{\eta\geq 0} c_\eta {\cal O}_\eta(x),\ x \geq 0.
\end{eqnarray}

However, the $\{{\cal E}_\eta(x)\}$ are dense with respect to $S_{\cal O}$, leading to an alternative representation:
\begin{eqnarray}
\Psi_{phys}(x) = \sqrt{2}\sum_{\eta \geq 0} d_\eta {\cal E}_\eta(x),\ x \geq 0.
\end{eqnarray}
 Thus, two very different looking representations span the same {\it physical} space. The first corresponds to the eigenstates of the HHO lying within $S_{\cal O}$. The other are the eigenfunctions of HHO lying within $S_{\cal E}$.
 Clearly this is inconsistent.

\subsubsection {Proof that $S_{\cal E}$ is dense with respect to $S_{\cal O}$ }

The following analysis is equivalent to the fact that the Associated (Generalized) Laguerre polynomials,
\begin{eqnarray}
L^{(\alpha)}_\eta(x^2) \propto \begin{cases} x^{-1}H_{2\eta+1}(x), \ \alpha = \frac{1}{2}, \\
 H_{2\eta}(x), \ \alpha = -\frac{1}{2}
\end{cases}
\end{eqnarray}
are complete on the positive real axis.

 Let ${\cal A}(x)$ correspond to any $L^2$ function on $\Re \equiv (-\infty,+\infty)$, we then have
\begin{eqnarray}
{\cal A}(x) =\sum_{n=0}^\infty  \frac{a_n}{\sqrt{2}}H_n(x) e^{-\frac{x^2}{2}}, 
\end{eqnarray}
or 
\begin{eqnarray}
{\cal A}(x) =\sum_{n=even}^\infty  \frac{a_n}{\sqrt{2}}H_n(x) e^{-\frac{x^2}{2}} + \sum_{n=odd}^\infty  \frac{a_n}{\sqrt{2}}H_n(x) e^{-\frac{x^2}{2}}, \cr
\end{eqnarray}
where $a_n = \frac{1}{\sqrt{2}}\int_{\Re}dx \ H_n(x)  {\cal A}(x)e^{-\frac{x^2}{2}}$.

Now assume that 
\begin{eqnarray}
{\cal A}(x) = 0,\ x \leq 0. 
\end{eqnarray}
It then follows that 
\begin{eqnarray}
0 = \sum_{n=even}^\infty a_n  H_n(-x) e^{-\frac{x^2}{2}}+\sum_{n=odd}^\infty a_n H_n(-x) e^{-\frac{x^2}{2}},
\end{eqnarray}
for $-x \leq 0$, 
or 
\begin{eqnarray}
\sum_{n=odd}^\infty a_n H_n(x) e^{-\frac{x^2}{2}}=\sum_{n=even}^\infty a_n H_n(x) e^{-\frac{x^2}{2}},
\end{eqnarray}
$x \geq 0$.
This in turn results in the completeness relations
\begin{eqnarray}
{\cal A}(x) & = &\sum_{\eta=0}^{N\rightarrow \infty} c_\eta H_{2\eta+1}(x) e^{-\frac{x^2}{2}}, \ x \geq 0;\\
& = &\sum_{\eta=0}^{N\rightarrow \infty} d_\eta H_{2\eta}(x) e^{-\frac{x^2}{2}}, \ x \geq 0,
\end{eqnarray}
for any $L^2$ configuration on the $\Re^{0,+}$ domain, that satisfies ${\cal A}(0) = 0$. The expansion coefficients (i.e. $\frac{2a_n} {\sqrt{2}}$), correspond to $c_\eta = \langle H_{2\eta+1}(x) e^{-\frac{x^2}{2}}|{\cal A}\rangle$ and
$d_\eta = \langle H_{2\eta}(x) e^{-\frac{x^2}{2}}|{\cal A}\rangle$. 

Given that the basis in Eq.(112) and Eq.(113) correspond to $\sqrt{2}{\cal O}_\eta(x)$ and $\sqrt{2}{\cal E}_\eta(x)$, respectively, we obtain Eq.(104) and Eq.(105).

From the above, we see that the odd states, 
${\cal O}_\eta(x) = \frac{1}{\sqrt{2}}H_{2\eta+1}(x) e^{-\frac{x^2}{2}}$ can be expressed through the appropriate infinite sum over the even states. In Fig. 12 and Fig. 13  we show this for the $\eta = 15$ odd state, ${\cal O}_{15}(x)$.

\begin{figure*}
\includegraphics{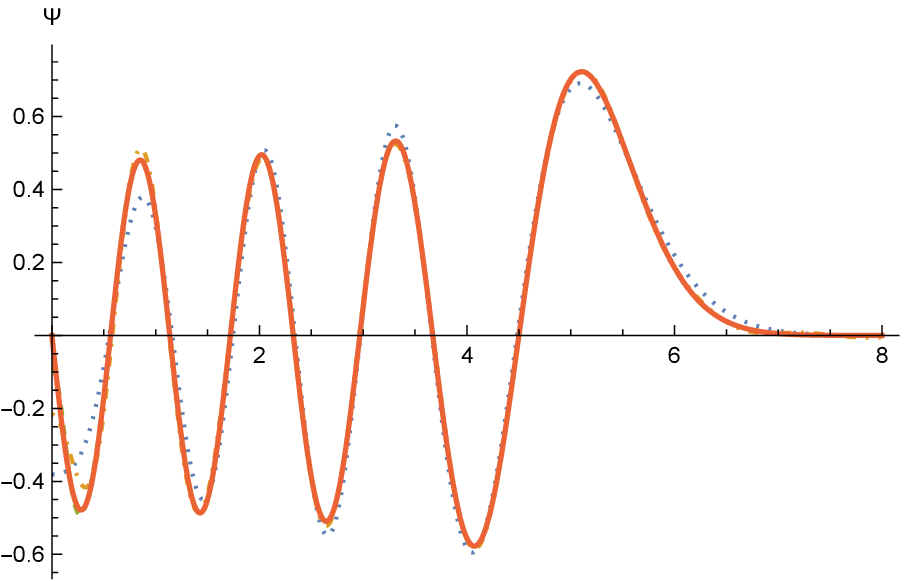}
\caption{ \label{fig:wide2} ${\cal O}_{15}(x) = \frac{1}{\sqrt{2}}H_{31}(x) e^{-\frac{x^2}{2}}$ generated from the ${\cal E}_\eta(x)$ expansion in Eq.(113): N = 10 (dots), 20 (dot-dash), 100 (dash), ${\cal O}_{15}(x)$ (solid) }
\end{figure*}

\begin{figure*}
\includegraphics{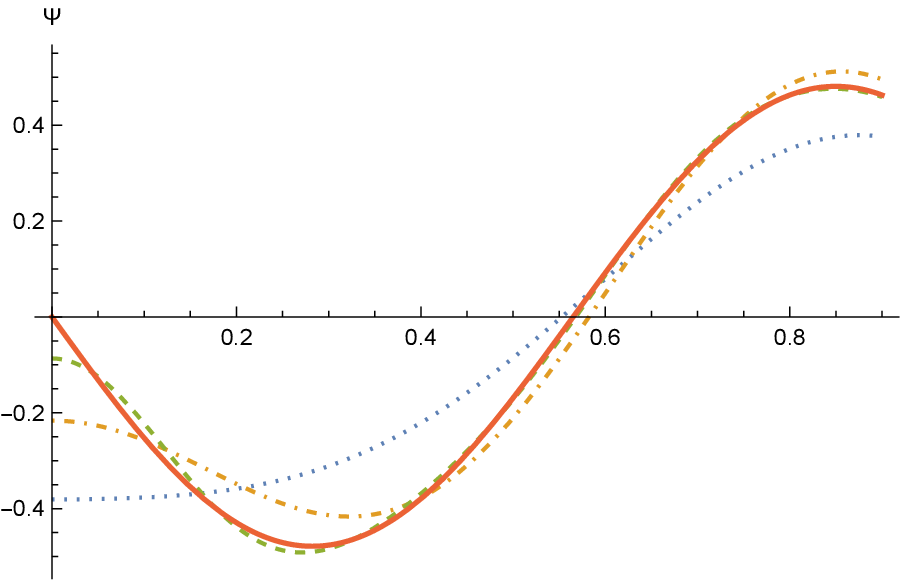}
\caption{ \label{fig:wide3} ${\cal O}_{15}(x) = \frac{1}{\sqrt{2}}H_{31}(x) e^{-\frac{x^2}{2}}$ generated from the ${\cal E}_\eta(x)$ expansion in Eq.(113): N = 10 (dots), 20 (dot-dash), 100 (dash), ${\cal O}_{15}(x)$ (solid) }
\end{figure*}

\subsection{The $-\partial_x^2\Psi +\frac{\gamma}{x^2}\Psi +x^2\Psi = 2E\Psi$ is an Exactly Solvable System for $\gamma > 0$}

Of interest is the fact that the system
\begin{eqnarray}
-\partial_x^2\Psi(x) +\frac{\gamma}{x^2}\Psi(x) +x^2\Psi(x) = 2E\Psi(x),
\end{eqnarray}
is exactly solvable. We want to prove this. Through substitution, we know that 
\begin{eqnarray}
\Psi_{gr}(x) = x^\alpha exp(-x^2/2),
\end{eqnarray}
provided
\begin{eqnarray}
\gamma = \alpha(\alpha-1).
\end{eqnarray}
The ground state energy is
\begin{eqnarray}
E_{gr} = \alpha+\frac{1}{2}.
\end{eqnarray}

In general, $\alpha$, as constrained by the previous relation is an indicial exponent, and the general (physical) solutions must behave, near the origin, as
\begin{eqnarray}
\Psi_{phys}(x) \sim x^\alpha A(x^2),
\end{eqnarray}
where $A$ is analytic in $x^2$. Furthermore, the Frobenius relation for the power series expansion coefficients of $A$ require $\alpha > 0$. From Eq.(116) we see that if $\alpha >1$, then the parameter $\gamma > 0$. 

As for the case examined earlier (i.e. Eq.(9)), we can work with
\begin{eqnarray}
\Phi(x) = x^\alpha exp(-x^2/2) \Psi(x).
\end{eqnarray}
Its differential equation becomes
\begin{align}
    \big( -2\alpha +(2\alpha-1)x^2 & -2Ex^2\big)\Phi(x) \nonumber\\
    & +\big(2\alpha x-2x^3\big)\partial_x\Phi(x) 
    -x^2\partial_x^2\Phi(x) = 0.
\end{align}
Its power moments, $\nu(p) = \int_0^\infty dx \ x^p\Phi(x)$. Since $\Phi(x) \sim O(x^{2\alpha})$, and $\alpha > 1$, we see that $\int_0^\infty x^p\partial_x\Phi(x) = -const \times (0^+)^{p+2\alpha}-p\nu(p-1) $; whereas
$\int_0^\infty x^p\partial_x^2\Phi(x) = -const \times {2\alpha}\ 0^{p+2\alpha-1} -p\Big(\int_0^\infty x^{p-1}\partial_x\Phi(x) = -const \times (0^+)^{p-1+2\alpha}-(p-1)\nu(p-2)\Big)$. 

The existence (i.e. integrability) of the moments requires $p+2\alpha > 0$, whereas no boundary terms contribute if $p+1+2\alpha > 0$. Both conditions are satisfied by $p > -2\alpha$. If we take $\alpha > 1$, then these conditions are satisfied (at least) by $p \geq -2$. The corresponding MER becomes
\begin{align}
\nu(p+2)\Big((2\alpha-1-2E) + & 2(p+3)\Big) \nonumber \\
& = \nu(p)\Big((2\alpha+ p+1)(p+2)\Big).
\end{align}
 The coefficient of $\nu(p)$ has a zero at $p = -2$.
Let us take $p = q-2$, where $q \geq 0$. We then have
\begin{align}
{\nu}(q)\Big(2\alpha+2q+1-2E  & \Big) \nonumber \\
& = \nu(q-2)\Big((2\alpha+ q-1)q\Big),
\end{align}
for $q \geq 0$.
We see that $2\alpha-1 > 0$. If $q = 0$, then $E_{gr} = \alpha+\frac{1}{2}$. Then, for all excited states we must have $\nu_{exc}(0) = 0$. We repeat the same logic as in the $\alpha = \frac{3}{2}$ case. Thus, the discrete state energies become (recall that $q = 0,2,4,\ldots,2n,\ldots$:

\begin{eqnarray}
E_n = \alpha+2n+\frac{1}{2}.
\end{eqnarray}
In terms of $\gamma =\{\frac{3}{4},2,\frac{15}{4},6\}$, we have
$\alpha =\frac{1}{2}\big(1+\sqrt{1+4\gamma})$, or 
$\alpha = \{\frac{3}{2}, 2,\frac{5}{2},3\}$.

The important observation is that the energy differences are the same, regardless of $\gamma > 0$, and $\alpha =\frac{1}{2}\big(1+\sqrt{1+4\gamma})$.

\subsection{OPPQ-AM Analysis of the (CQ) Translated {\it Walled Harmonic Oscillator}}

Consider the canonical quantization version of the harmonic oscillator with an infinite potential barrier at $x = -b$ (i.e. the {\it walled harmonic oscillator})
\begin{eqnarray}
-\partial_x^2\Psi(x) + x^2\Psi(x) =2E\Psi(x),
\end{eqnarray}
for $x \in (-b,+\infty) \equiv {\cal D}_b$. Since the problem is defined on the entire real axis, the implicit infinite barrier (wall), forces the wavefunction to satisfy  $\Psi(x) = 0$, for $x \leq -b$.  

Since the parameter `$b$' characterizes the {\it wall} potential, and $E$ is arbitrary, we will denote the solutions to Eq.(125) by $\Psi_{b;E}(x)$.

For any $E > 0$ there will be square integrable solutions on ${\cal D}_b$, satisfying
\begin{eqnarray}
\Psi_{b;E}(x) & \sim & Poly(x)\ exp(-{\frac {1}{2}}x^2), \ x \rightarrow \infty \cr
\Psi_{b;E}(-b) & = & finite, \cr
\partial_x\Psi_{b;E}(-b)& = & finite.
\end{eqnarray}

The physical solutions obey 
\begin{eqnarray}
\Psi_{b:E_n^{(p)}}(-b) = 0;
\end{eqnarray}
whereas unphysical solutions
satisfy
\begin{eqnarray}
\partial_x\Psi_{b;E_n^{(u)}}(-b) = 0,
\end{eqnarray}
where '$n$' is the quantum number, and we have made explicit the fact that $E_n^{(p)}\neq E_n^{(u)}$.

We denote the wavefunctions and energies for the full harmonic oscillator problem by
$\Psi_{\infty;n}(x)$ and $E_{\infty;n} = n+\frac{1}{2}$, respectively, where `$n$' is the quantum number.

We know that at position $x = -b$, for  $b >> 1$,  both $\Psi_{\infty;n}(-b)$  and $\partial_x\Psi_{\infty;n}(-b)$
will become exponentially small (assuming unit probability density normalization):
\begin{eqnarray}
|\Psi_{\infty;n}(-b)| \sim Poly(x) \ exp(-\frac{b^2}{2}) << 1, \\
|\partial_x\Psi_{\infty;n}(-b)| \sim Poly(x) \ exp(-\frac{b^2}{2}) << 1.
\end{eqnarray}

Although we know that Eq.(126) must have acceptable solutions, it is instructive to understand how they develop.
Let $E = E_{\infty;n}+\delta E$. As $\delta E$ varies, the corresponding $L^2({\cal D}_b,dx)$ solutions will satisfy Eq.(125). At special, very small values of $\delta E \approx 0$, we will find the energy values satisfying Eq.(126) and Eq.(127).
\subsubsection{The Associated MER}

As before, we work in transformed coordinate $\chi = x+b > 0$. The new system is
\begin{eqnarray}
 -\partial_\chi^2\Psi(\chi) + (\chi^2 +\beta\chi)\Psi(\chi) = {\cal E}\Psi(\chi),
\end{eqnarray}
where $\beta = -2b$ and ${\cal E} = 2E-b^2$.

The Stieltjes moments $v(p) = \int_0^\infty d\chi \chi^p \Psi(\chi)$
satisfy the relation 
\begin{align}
 v(p+2) = -\beta v(p+1) + {\cal E}v(p) + p(p-1)v(p-2) \nonumber\\ + \delta_{p,1}\Psi(0) - \delta_{p,0} \Psi'(0),
\end{align}
$p \geq 0$.
For the physical and unphysical solutions in Eq.(126) and Eq.(127) we obtain:

\begin{align}
 v_{phys}(p+2) = & -\beta v_{phys}(p+1) + {\cal E}v_{phys}(p)  \nonumber\\ &   + p(p-1)v_{phys}(p-2)   - \delta_{p,0} \Psi'(0),
\end{align}

\begin{align}
 v_{unphys}(p+2) = & -\beta v_{unphys}(p+1) + {\cal E}v_{unphys}(p)  \nonumber\\ &   +  p(p-1)v_{unphys}(p-2) + \delta_{p,1} \Psi(0),
\end{align}
 $p \geq 0$.
 These correspond to an (effectively) $m_s =2$  MER relation, due to the boundary term. That is $1+m_s = 3$ linear (homogeneous) initialization variables are required to generate the MER relation: $\{\Psi(0) \ {\rm or} \ \Psi'(0), v(0),v(1)\}$.
 
 In either case, we can take $\ell \in \{-1,0,1\}$, where $v_\ell = \{v(0),v(1)\}$, if $\ell = 0,1$, respectively; and $v_{-1} = \Psi'(0) $, or $v_{-1} = \Psi(0) $, for the boundary term.
 
 We then have
 
 \begin{align}
 M_{phys;E} & (p+2,\ell)   \nonumber \\ & = \ 2b M_{phys;E}(p+1,\ell) + (2E-b^2)M_{phys;E}(p,\ell)  \nonumber\\ &   + p(p-1)M_{phys;E}(p-2,\ell)   -  \delta_{p,0}\delta_{\ell,-1},
\end{align}
where 
\begin{eqnarray}
M_E(\ell_1,\ell_2) = \begin{cases} \delta_{\ell_1,\ell_2},\  0 \leq \ell_{1,2} \leq 1,  \\
0,\  0\leq \ell_1 \leq 1, \ell_2 = -1,\\
\delta_{\ell_2,-1}, \ell_1 = -1.
\end{cases}
\end{eqnarray}

For the unphysical solutions we have
 \begin{align}
 M_{unphys;E} & (p+2,\ell)  \nonumber \\
 =\ & 2b M_{unphys;E}(p+1,\ell) + (2E-b^2)M_{unphys;E}(p,\ell)  \nonumber\\ &   + p(p-1)M_{unphys;E}(p-2,\ell)   +  \delta_{p,1}\delta_{\ell,-1},
\end{align}
$p \geq 0$, and the same initialization conditions in Eq.(135).

We note that  \begin{eqnarray}
v(p) = \sum_{\ell = -1}^1 M_E(p,\ell) v(\ell),
\end{eqnarray}
for $ p \geq 0$.

We will use the   weight $R(\chi) = exp(-\frac{1}{2}\chi^2+b \chi)$, which captures the asymptotic form of the $L^2({\cal D}_b,dx)$ solutions (i.e. $exp\big(-\frac{1}{2}(\chi-b)^2\big)$, for $0 \leq \chi < \infty$, and $b \rightarrow \infty$). Its differential equation $R'(\chi) = (b-\chi)R(\chi)$ results in the Stieltjes moment equation by which the orthonormal polynomials are generated: $w(p+1) = \delta_{p,0} +b w(p)+pw(p-1)$, for $p \geq 0$.

If we denote the orthonormal polynomials by $P_n(\chi) = \sum_{j=0}^n\Xi_j^{(n)}x^j$, we obtain the desired $c_n = \sum_{\ell = -1}^1 \Lambda_\ell^{(n)}(E)v_\ell$, where

\begin{eqnarray}
\Lambda_\ell^{(n)}(E) = \sum_{j =0}^n\Xi^{(n)}_jM_E(j,\ell).
\end{eqnarray}

Within the OPPQ-AM formulation, no normalization of the initialization expression $\{v_\ell|-1\leq \ell \leq 1\}$ are required. Within the OPPQ-BM, we can pick a unit vector normalization.

The OPPQ-AM quantization conditions become
\begin{eqnarray}
Det\Big(\Lambda^{(N-\ell_1)}_{\ell_2}(E)\Big) = 0,
\end{eqnarray}
for $-1 \leq \ell_{1,2}\leq 1$, and $N \rightarrow \infty$.

Both the OPPQ-AM energy approximants, and the OPPQ-BM energy estimates (i.e. the local minima of the corresponding eigenvalue energy functions, $\partial_E\lambda_N(E) = 0$) concur. These are quoted in Tables 8-11. We note that as a check on the results, the eigenstates of the full harmonic oscillator have zero-nodes at known positions. If $H_n(b) = 0 $, then our analysis should yield the correct eigenenergy state $E_n = n+\frac{1}{2}$. The same applies for the zero derivatives $\partial_x\big( H_n(b) exp(-b^2/2)\big) = 0$. These are confirmed in Tables 10-11.

We see that  the $\Psi(-b) = 0$ are the only physical states within the canonical quantization of the walled harmonic oscillator (i.e. defined on the entire real axis). These become the full states of the full harmonic oscillator when $b \rightarrow \infty$. The unphysical states (i.e. $\partial_x\Psi(-b) = 0$) also behave similarly.

\begin{table*}
\caption{\label{}  Walled Harmonic Oscillator:$-\Psi''(x)+x^2\Psi(x) =2E\Psi(x)$, $x \in (-b,\infty)$, $\Psi(-b) = 0$,  $N = 100$ (i.e. Eq.(139)).
}
\begin{ruledtabular}
\begin{tabular}{llllll}
\hline
$b$& $E_{0}$ & $E_{1}$ &$E_{2}$ & $E_{3}$  & $E_{4}$ \\
\hline
0.0&   1.5&     3.5 & 5.5 &    7.5    &   9.5   \\
0.5&  1.030383 &	2.752738 &	4.542103&	6.366152 &	8.212066 \\
1.0 &	0.7342339	&2.197463	&3.780191	&5.429814 &	7.122444\\
1.5	&0.5787399 &	1.823134	&3.208399	&4.686630	&6.227426\\
2.0&	0.5178817&	1.611507	&2.820444	&4.133244&	5.524367\\
2.5&	0.5024589&	1.524564&	2.603632&	3.766711&	5.012528\\
3.&	0.5001954&	1.503020	&2.519911&	3.575075&	4.691235\\
3.5&	0.5000090&	1.500200&	2.501977&	3.511446&	4.543240\\
4.0&	0.5000002 &	1.500007	&2.500101 &	3.500844&	4.504784\\
4.5&	0.5&	1.5	&2.500003&	3.500031&	4.500246\\
5.0	&0.5 &	1.5	&2.5&	3.500001&	4.500006\\
10.0&0.5 & 1.5 &2.5 & 3.5& 4.5\\
\hline
\end{tabular}
\end{ruledtabular}
\end{table*}

\begin{table*}
\caption{\label{}  Walled Harmonic Oscillator:$-\Psi''(x)+x^2\Psi(x) =2E\Psi(x)$, $x \in (-b,\infty)$, $\partial_x\Psi(-b) = 0$,  $N = 100$ (i.e. Eq.(139)).
}
\begin{ruledtabular}
\begin{tabular}{llllll}
\hline
$b$ & $E_{0}$ & $E_{1}$ &$E_{2}$ & $E_{3}$  & $E_{4}$ \\
\hline
0.0 &	0.5 &	2.5 &	4.5	& 6.5 &	8.5\\
0.5	& 0.3177644 &	1.897087 &	3.649086 &	5.454896 &	7.289546\\
1.0 &	0.3094597 &	1.5 &	2.998601 &	4.609231 &	6.278425\\
1.5	& 0.4007468	& 1.324338	& 2.552000 &	3.961911 &	5.464325\\
2.0 &	0.4757094 &	1.367518 &	2.335409 &	3.521749 &	4.849426\\
2.5 &	0.496951 &	1.466184 &	2.372225 &	3.328418 &	4.451125\\
3.0 &	0.4997757 &	1.496261 &	2.472762 &	3.399329 &	4.328252\\
3.5	& 0.4999901	& 1.499773	& 2.497608 &	3.484842 &	4.439140\\
4.0&	0.4999997 &	1.499992 &	2.499887 &	3.499012 &	4.493994\\
4.5 &	0.5	& 1.5	& 2.499997 &	3.499966	& 4.499721 \\
5.0 &	0.5	& 1.5& 2.5 &	3.499999 &	4.499993\\
10.0 &	0.5 &	1.5 &	2.5 &	3.5 &	4.5\\
\hline
\end{tabular}
\end{ruledtabular}
\end{table*}

\begin{table*}
\caption{\label{}  Zeroes for 
${\cal H}(x) = 0$ and $\partial_x{\cal H}(x) = 0$,
where ${\cal H}(x) = H_{10}(x) exp(-x^2/2)$.
}
\begin{ruledtabular}
\begin{tabular}{ccccccc}
\hline
 & $x_1$&$x_2$&$x_3$&$x_4$&$x_5$&$x_6$\\
 \hline
${\cal H}_{10}(x) = 0$ &$\pm 0.3429013272$ &$ \pm 1.0366108298$&
$\pm 1.7566836493$ & $\pm 2.5327316742$ &
$\pm 3.4361591188$& \\
\hline
${\cal H}_{10}'(x) = 0$ & $ \hspace{7pt} 0$ &
 $\pm 0.6885543048$&
$\pm 1.3938231562$ & $\pm 2.1388620065$ &
$\pm 2.9695588950$ & $\pm 4.0853566875$\\
\end{tabular}
\end{ruledtabular}
\end{table*}

\begin{table*}
\caption{\label{} Comparison of OPPQ-AM (N=100) Energies ($0< E \leq 12$) for $\Psi(-b) = 0$ and $\partial \Psi_x(-b) = 0$,  and the zeroes in Table X.
}
\begin{ruledtabular}
\begin{tabular}{lllllllllllll}
\hline
$b$ & $E_{0}$ & $E_{1}$ &$E_{2}$ & $E_{3}$  & $E_{4}$ &$E_{5}$ & $E_{6}$ &$E_{7}$ & $E_{8}$  & $E_{9}$&$E_{10}$ &$E_{11}$\\
\hline
  0.3429013272 &1.15805&	2.96640&	4.82168&	6.70090&	8.59518&
  {\bf 10.5}& & & & & & \\
  1.0366108298	& 	0.71856 &	2.16407&	3.73194&	5.36889&	7.05033&	8.76330&	{\bf 10.5}	& & & & &\\

  1.7566836493 &	0.53894 &	1.69592	&2.98677 &	4.37899&	5.84262	&7.35823&	8.91344	&{\bf 10.5}& & & &\\

  2.5327316742	&  0.50212
 &1.52182
 &2.59471
& 3.74908
& 4.98570
& 6.29138
 &7.65237
& 9.05790
& {\bf 10.5}
& 11.97274
& &\\
  3.4361591188 &0.50001
 &	1.50029
 &  2.50276
 &	3.51514
 &	4.55422
 &  5.63893
 &  6.77896
 &  7.97326
&   9.21584
&  {\bf 10.5}
&  11.81988
& \\
  \hline
0 &	0.5 &	2.5 &	4.5 &	6.5 &	8.5 & {\bf 10.5}& & & & & &\\
0.6885543048 &	0.29696 &	1.72280&	3.38031&	5.11272&	6.88500&	8.68319&	{\bf 10.5}& & & & &\\
1.3938231562&	0.37889 &	1.34181 &	2.62903&	4.08261&	5.62073&	7.21224&	8.84168&	{\bf 10.5}& & & & \\
2.1388620065&	0.48552&	1.39942&	2.32255&	3.44041&	4.71559&	6.08364 &	7.51347&	8.98887&	{\bf 10.5}& & & \\
2.9695588950 &	0.49973&	1.49565&	2.46916&	3.39139&	4.32663&	5.38701&	6.56624&	7.82532&	9.14092&	{\bf 10.5}&	11.89439 &\\
 4.0853566875& 0.50000&	1.50000&	2.49994&	3.49942&	4.49627&	5.48253&	6.44098&	7.36975&	8.32871&	9.37397&	{\bf 10.5}&	11.68532 \\

\hline
\end{tabular}
\end{ruledtabular}
\end{table*}

\section{References}
\bibliography{aipsamp}

\noindent 1. R. Fantoni and J. R. Klauder, {\it Affine quantization of $(\Phi^4)_4$
 succeeds while canonical quantization fails},
Phys. Rev. D {\bf 103}, 076013 (2021).\\

\noindent 2. R. Fantoni and J. R. Klauder, {\it Monte Carlo evaluation of the continuum limit of the two point function of two Eucldiean Higgs real scalar fields subject to affine quantization}, Phys. Rev. D {\bf 104}, 054514 (2021).\\

\noindent 3. Klauder J R 2020, {\it Quantum gravity made easy}, Journal of High Energy
Physics, Gravitation and Cosmology, {\bf 6}, 90-102.\\

\noindent 4. J. R. Klauder, {\it The benefits of affine quantization}, J. High
Energy Physics, Gravitation Cosmol. {\bf 6}, 175-185 (2020).\\

\noindent 5. E. Frion and C. R. Almeida, {\it Affine quantization of the Brans-Dicke theory: Smooth bouncing and the equivalence between the Einstein and Jordan frames} Phys. Rev. D {\bf 99}, 023524 (2019)\\

\noindent 6. H. Bergeron, A. Dapor, J. P. Gazeau, and P. Malkiewicz, {\it Smooth big bounce from affine quantization}, Phys. Rev. D {\bf 89} 083522 (2014).\\

\noindent 7. M. Fanuel and S. Zonetti, {\it Affine quantization and the initial cosmological singularity}, EPL {\bf 101}, 10001 (2013).\\

\noindent 8. J. R. Klauder, {\it Recent results regarding affine quantum gravity}, J. Math. Phys. {\bf 53}, 082501 (2012).\\

\noindent 9. L. Gouba, {\it Affine quantization on the half line}, J. High Energy Physics, Gravitation and Cosmol., J. High Energy Physics, Gravitation Cosmol. {\bf 7}, 352-365 (2021).\\

\noindent 10. J. A. Shohat and J. D.  Tamarkin J D,{\it The Problem of Moments} (American Mathematical Society, Providence, RI, 1963).\\ 

\noindent 11. C. R. Handy and D. Bessis, {\it Rapidly Convergent Lower Bounds for the Schrodinger Equation Ground State Energy},  Phys. Rev. Lett. {\bf 55}, 931 (1985). \\

\noindent 12. C. R. Handy, D. Bessis, G. Sigismondi, and T. D. Morley, {\it Rapidly Converging Bounds for the Ground State Energy of Hydrogenic Atoms in Superstrong Magnetic Fields}, Phys. Rev. Lett., {\bf 60} 253-256 (1988).\\

\noindent 13. C. R. Handy, D. Bessis, and T. D. Morley, {\it Generating quantum energy bounds by the moment method: A linear programming approach},  Phys. Rev. A, {\bf 37} 4557 -4569 (1988). \\

\noindent 14. C. R. Handy {\it Moment Method Quantization of a Linear Differential Eigenvalue Equation for $|\Psi|^2$}, Phys. Rev. A {\bf 36}, 4411 (1987).\\

\noindent 15. C. R. Handy, {\it Nonnegativity and Moment Quantization for $|\Psi|^2 $}, Phys. Lett. {\bf A 124}, 308 (1987).\\

\noindent 16. C. R. Handy, {\it Generating converging bounds to the (complex) discrete states of the $P^2+iX^3+i\alpha X$ Hamiltonian}, J. Phys. A: Math. Gen., {\bf 34} (2001). \\

\noindent 17. C. R. Handy, {\it Exact Christoffel-Darboux expansions: A new multidimensional, algebraic, eigenenergy bounding method}, Physica Scripta, {\bf 96},   075201 (2021).\\

\noindent 18. C. R. Handy and D. Vrinceanu, {\it Orthogonal polynomial projection quantization: a new Hill determinant method}, J. Phys. A: Math. Theor. ,  {\bf 46}  135202 (2013).\\

\noindent 19. C. R. Handy and D. Vrinceanu, {\it Rapidly converging bound state eigenenergies for the two dimensional quantum dipole},  J. Phys. B: At. Mol. Opt. Phys., {\bf 46},  115002 (2013).\\ 

\noindent 20. C. M. Bender and S. A. Orszag {\it Advanced Mathematical Methods for Scientists and Engineers} (Springer-Verlag, New York, 1999)

\noindent 21. C. R. Handy, {\it Singular perturbation-strong coupling field theory and the moments problem} {\em Phys. Rev. D},  {\bf 24}, 378-383 (1981). \\

\noindent 22. A. Grossmann and J. Morlet, {\it Decomposition of Hardy Functions into Square Integrable Wavelets of Constant Shape}, SIAM J. Math. Anal., {\bf 15} 723-736 (1984). \\

\noindent 23. I. Daubechies {\it Ten Lectures on Wavelets} (SIAM, 1992).\\

\noindent 24. Handy C R and Murenzi R, {\it Moment-Wavelet Quantization: a first principles analysis of quantum mechanics through continuous wavelet transform theory}, Phys. Lett. A {\bf 248}, 7-15 (1998).

\noindent 25. R. L. Hall, {\it Spiked Harmonic Oscillators}, J. Math. Phys. {\bf 43}, 94 (2002).\\

\noindent 26. A. F. Nikiforov and V. B. Uvarov, {\em{Special functions of Mathematical Physics}} (Birkhauser, Boston, 1988)\\

\noindent 27. F. Cooper, A. Khare, and U. Sukhatme, {\it Supersymmetry and quantum mechanics}, Phys. Rep. Elsevier {\bf 251},  267-385 (1995).\\

\noindent 28. S. Boyd and L. Vandenberghe {\it Convex Optimization} (New York: Cambridge University Press, 2004).\\

\noindent 29. J. B. Lasserre, {\it { Moments, Positive Polynomials and Their Applications}} (Imperial College Press, London, 2010).\\

\noindent 30. V. Chvatal, {\it Linear Programming} (Freeman, New York, 1983).\\

\noindent 31.Y. P. Kravchenko, M. A. Liberman,  and B.  Johansson B, {\it Exact solution for a hydrogen atom in a magnetic field of arbitrary strength}, Phys. Rev. A {\bf 54}, 287 – 305 (1996).\\

\noindent [32] C. Schimerczek  and G. Wunner {\it Accurate 2d finite element calculations for hydrogen in magnetic fields of arbitrary strength},  Comp. Phys. Comm. {\bf 185}, 614-621 (2014).

\end{document}